\documentclass{aa}    

\usepackage{upgreek}
\usepackage{color}
\usepackage{hhline}
\usepackage{longtable}
\usepackage{graphicx}
\usepackage{multicol}

\usepackage[varg]{txfonts}
\usepackage{upgreek} 
\usepackage{multirow} 
\usepackage[FIGTOPCAP, center, nooneline]{subfigure}
\usepackage[]{natbib}

\usepackage[varg]{txfonts}
\usepackage{microtype}
\usepackage{hyperref}
\usepackage{booktabs}
\usepackage{xcolor}
\usepackage{hyperref}
\usepackage{graphicx}
\usepackage{multirow}
\usepackage{float}

\hypersetup{%
  colorlinks=true,
  linkcolor=blue,
  citecolor=blue
}

\usepackage{etoolbox}

\let\oldcitet=\citet

\renewcommand{\citet}[1]{\textcolor[rgb]{0,0,1}{\oldcitet{#1}}}

\begin{document}

\title{The role of highly vibrationally excited H$_2$ initiating the N chemistry}

\subtitle{Quantum study and 3$\sigma$ detection of NH emission in the Orion Bar PDR}

\titlerunning{Formation of NH from highly vibrationally excited H$_2$ in PDRs} 
\authorrunning{Goicoechea \& Roncero}

 \author{Javier R.\,Goicoechea\inst{1}
          \and
        Octavio Roncero\inst{1}}

\institute{Instituto de F\'{\i}sica Fundamental (IFF),
     CSIC. Calle Serrano 121-123, 28006, Madrid, Spain. \email{javier.r.goicoechea@csic.es}
}

   \date{Received 28 April 2022 / Accepted 9 June 2022}



\abstract{The formation of  hydrides by gas-phase reactions between H$_2$  and a heavy element atom is a very selective process. 
Reactions with ground-state  neutral carbon, oxygen, nitrogen, and  sulfur atoms are very endoergic and have high energy barriers because
the H$_2$ molecule has to be fragmented before a hydride bond is formed. 
In cold interstellar clouds, these barriers
exclude the formation of CH, OH, NH, and SH radicals 
through hydrogen abstraction reactions.
Here we study a very energetically unfavorable process, the reaction of
\mbox{N\,($^4S$)} atoms with H$_2$ molecules. We calculated
the reaction rate coefficient for H$_2$ in different vibrational levels, using quantum methods  for \mbox{$v$\,=\,0$-$7} and quasi-classical methods up to  \mbox{$v$\,=\,12;}
for comparison purposes, we also calculated the 
rate coefficients of the analogous  reaction \mbox{S\,($^3P$)+\,H$_2$($v$)\,$\rightarrow$\,SH\,+\,H}. Owing to the high energy barrier, these rate coefficients  increase
with $v$ and also with the gas temperature. We implemented the new rates in the Meudon 
photodissociation region (PDR) code and studied their effect on models with different ultraviolet (UV) illumination conditions.
In strongly UV-irradiated dense gas (Orion Bar conditions),
the presence  of H$_2$ in highly vibrationally excited
levels ($v$\,$\geq$\,7) enhances the NH abundance by two orders of magnitude (at the PDR surface) compared to models that use the thermal rate coefficient
for reaction \mbox{N($^4S$)\,$+$\,H$_2$\,$\rightarrow$\,NH + H}. The increase in
NH column density, $N$(NH),  across the PDR is a factor of $\sim$25. We investigate the excitation and detectability of submillimeter NH rotational emission  lines. 
 Being a hydride, NH excitation is very subthermal ($T_{\rm rot}$\,$\ll$\,$T_{\rm k}$) even in warm and dense  gas.  We explore existing Herschel/HIFI observations of the Orion Bar and Horsehead PDRs. We report a 3$\sigma$ emission feature at the $\sim$974\,GHz frequency of the NH \mbox{$N_J$\,=\,$1_2-0_1$} line  toward the  Bar. The emission level  implies \mbox{$N$(NH)\,$\simeq$\,10$^{13}$\,cm$^{-2}$}, which is consistent 
 with PDR  models using the new  rate coefficients for  reactions between N 
 and  \mbox{UV-pumped H$_2$}. This formation  route dominates over hydrogenation reactions involving the less abundant N$^+$ ion.
 JWST observations will quantify the amount and reactivity of
UV-pumped H$_2$  in many interstellar and circumstellar  environments.}

\maketitle

\section{Introduction}\label{sec:introduction}

The abundances of interstellar hydrides provide key information
about the physical conditions where they are found, the H$_2$/H  \mbox{fraction}, and the ionization rate  \mbox{\citep{Gerin16}}.
Much of the interstellar chemistry begins with the reaction of H$_2$ molecules with
a heavy element atom X. The dominant ionization state of  X in neutral clouds depends on its ionization potential (IP). Indeed, only  far-ultraviolet (FUV) photons with energies below
13.6\,eV \mbox{-- that is to say those that} cannot ionize hydrogen --  penetrate  diffuse neutral
clouds
\citep[][]{Snow06}  and the illuminated rims of denser molecular clouds, in other words their photodissociation regions \citep[PDRs;][]{Hollenbach97,Wolfire22}.
In consequence, the initial gas reservoir
is \mbox{ionized} for elements such as   
C (IP\,=\,11.26\,eV) and S (10.36\,eV), but
neutral\footnote{The existence of small amounts of O$^+$ and N$^+$ ions in these clouds is related to the ionization produced by energetic cosmic-ray particles and \mbox{X-rays}. The presence of abundant C$^+$ and S$^+$ ions is generally related to the presence of stellar or secondary FUV photons.}  for elements such as N (IP\,=\,14.53\,eV) and O (13.62\,eV).

Even for small amounts of  H$_2$ in the gas, the reaction
\begin{equation}
\rm X\,+\,H_2\,\rightarrow\,XH\,+\,H, 
\label{reaction-general}
\end{equation}
with 
\mbox{X\,=\,X$^0$} or X$^+$ and \mbox{XH\,=\,XH$^0$ or XH$^+$}, initiates  
the formation of interstellar hydrides.
This happens in diffuse  clouds of initially atomic gas \citep[\mbox{$n_{\rm H}$\,$\simeq$\,100$-$500\,cm$^{-3}$}; e.g.,][]{Federman95,Gerin10,Neufeld10,Sonnentrucker10,Schilke14}, 
 and also in  dense PDRs, as demonstrated by the detection of
OH, CH, CH$^+$, SH$^+$, OH$^+$, and HF rotational emission lines toward the Orion Bar \citep[][]{Naylor10,Goicoechea11,Goicoechea21b,Nagy13,Tak13,Joblin18,Kavak19}.

 The rate coefficient  of reaction~(\ref{reaction-general})
greatly depends on the element X involved (see Sect.~\ref{sec:reactivity_general}).
If we restrict ourselves to neutral C, N, O, and S, these
hydrogen abstraction reactions are very endoergic and have substantial energy barriers.  When X is a cation, reaction~(\ref{reaction-general}) are endoergic but barrierless.
The only exception is \mbox{X\,=\,O$^+$}, for which the reaction is exothermic and thus fast. 
Indeed, OH$^+$ is readily detected in interstellar clouds 
\mbox{\citep[e.g.,][]{Indriolo15}}.
In general, the endothermicity of these reactions 
(\mbox{$\Delta$$E$\,/\,$k_{\rm B}$} of several \mbox{thousand} kelvin) is significantly above the bulk gas temperatures in diffuse   interstellar clouds (\mbox{$T$\,$\simeq$\,30$-$100\,K}).
 In particular, the presence of  CH$^+$ and SH$^+$ absorption lines toward
these  low density 
 clouds has always been
  puzzling \citep[e.g.,][]{Godard12}. 
Plausible explanations are the formation of
 these  hydrides  in hot gas ($T$\,$\simeq$\,1000\,K) heated by 
shocks \citep[e.g.,][]{Elitzur80, Pineau86, Draine86, Neufeld02}, dissipation of  turbulence and ion-neutral drift \citep[][]{Godard14}, or triggering by the presence
of   hot H$_2$ from phase mixing \citep{Lesaffre07,Valdivia17}. 
These mechanisms, however, are not 
relevant in dense molecular clouds and PDRs.
 
 \mbox{Reaction~(\ref{reaction-general})} for  \mbox{X\,=\,N} is 
a very energetically unfavorable process: \mbox{$\Delta$$E$\,/\,$k_{\rm B}$\,$\simeq$\,15,000\,K}, and it  has a high barrier. Hence, this reaction is not considered to be
a relevant formation pathway for NH in diffuse  clouds \citep[][]{Godard10}, 
despite line absorption observations showing the presence of NH  \citep{Meyer1991,Crawford97,Persson10,Persson12}.

It has long been suggested that the presence of high densities of \mbox{FUV-pumped} H$_2$  in  dense PDRs  enhances the reactivity of  
\mbox{reaction~(\ref{reaction-general})} 
 \citep[][]{Stecher72,Freeman82,Tielens85a,Sternberg95}.
The prototypical example is the Orion Bar, a nearly \mbox{edge-on} rim of the Orion cloud \citep[][]{Tielens93,Goico16}.  \mbox{FUV-pumping} in the \mbox{Lyman} and \mbox{Werner} bands of H$_2$, followed by radiative or collisional de-excitation, 
populates H$_2$ in highly vibrationally excited levels
within the ground-electronic state
\citep[][]{Black76,Sternberg89,Burton90}. Infrared ro-vibrational  lines from these very excited levels are readily detected
in PDRs \citep[][]{Kaplan21}.

 In order to treat the increase in
reactivity of vibrationally excited H$_2$, early PDR models assumed state-specific rate coefficients in which the energy $E_v$ of each H$_2$ vibrational state ($v$) was
subtracted from the reaction endothermicity. That is, 
\mbox{$k_v(T)$\,$\propto$\,exp\,$(-[\Delta E - E_v]/k_{\rm B}T)$}. However, the reactivity of 
\mbox{reaction~(\ref{reaction-general})} for H$_2(v\geq1)$ is a very selective process, and the above assumption
is often not realistic and can lead to uncertain predictions. 
 Hence, an accurate estimation of hydride  abundances   requires one to compute the H$_2$($v$) level populations
 and to implement \mbox{$v$-state-specific} rates in chemical networks \citep{Agundez10}. This also applies to the chemistry of irradiated
 protoplanetary disks \citep[e.g.,][]{Fedele13,Ruaud21}. 
 
When laboratory measurements are not possible, state-specific reaction rate coefficients, $k_v(T)$, can be determined through 
ab initio calculations of the potential energy surface (PES) followed by a  study of the reaction dynamics.
Detailed calculations of state-specific rates of \mbox{reaction~(\ref{reaction-general})} 
exist for 
X\,=\,C$^+$ \citep{Zanchet13b}, X\,=\,S$^+$ \citep{Zanchet13a,Zanchet19},
X\,=\,O$^+$ \citep{Gomez-Carrasco-etal:14},
X\,=\,N$^+$, \citep{Grozdanov_16,Gomez-Carrasco_22},
 X\,=\,S and SH$^+$ \citep{Goicoechea21b}, and  X\,=\,O \citep{Veselinova21}. The above calculations contributed to explain the 
 observed abundances of hydride molecules in interstellar and circumstellar environments. Furthermore,
 the formation of very reactive \mbox{hydrides} (most notably CH$^+$) from hydrogen abstraction reactions  involving \mbox{vibrationally} excited H$_2$ also determines much of the hydride excitation \citep[][]{Godard13,Faure17} and explains the observed  extended
 CH$^+$  emission from \mbox{FUV-irradiated} molecular cloud surfaces \citep{Morris16,Goicoechea19}.

Here we study the state-specific behavior  of reaction
\begin{equation}
{\rm N\,(^4\it{S})\,+\,{\rm H_2\,(^1\Sigma_{g}^{+},\,\it{v}={\rm{0-12}})}\,\rightarrow\,{\rm NH\,(^3\Sigma^{-})\,+\,H}},
\label{reac-NH}
\end{equation}    
through dynamical (quantum and quasi-classical) calculations. Based on PDR model predictions using our new rate
coefficients, we also searched for NH  emission lines  toward 
the Orion Bar and the Horsehead PDRs.

The paper is organized as follows. In Sec.~\ref{sec:reactivity_general} we briefly discuss the most salient differences of 
reaction~(\ref{reaction-general}) when X\,=\,C, N, O, and S are neutral or ionized.
In Sec.~\ref{sec:N-reaction} we focus on reaction~(\ref{reac-NH}) and summarize the details of our dynamical calculations 
to determine the state-specific reactive rate coefficients.
In Sec.~\ref{sec:PDR_models} we investigate their impact on PDR models
adapted to the physical and illumination conditions in the two prototypical PDRs.
Finally, in Sec.~\ref{sec:Detectability} we study the  excitation of NH rotational lines, search for  NH emission lines in submillimeter observations, and present a 3$\sigma$ detection toward the Orion Bar PDR.

\section{Reactivity of H$_2$\,($v\geq 1$) with abundant elements}\label{sec:reactivity_general}

\begin{figure}[t]
\centering    
\includegraphics[scale=0.46, angle=0]{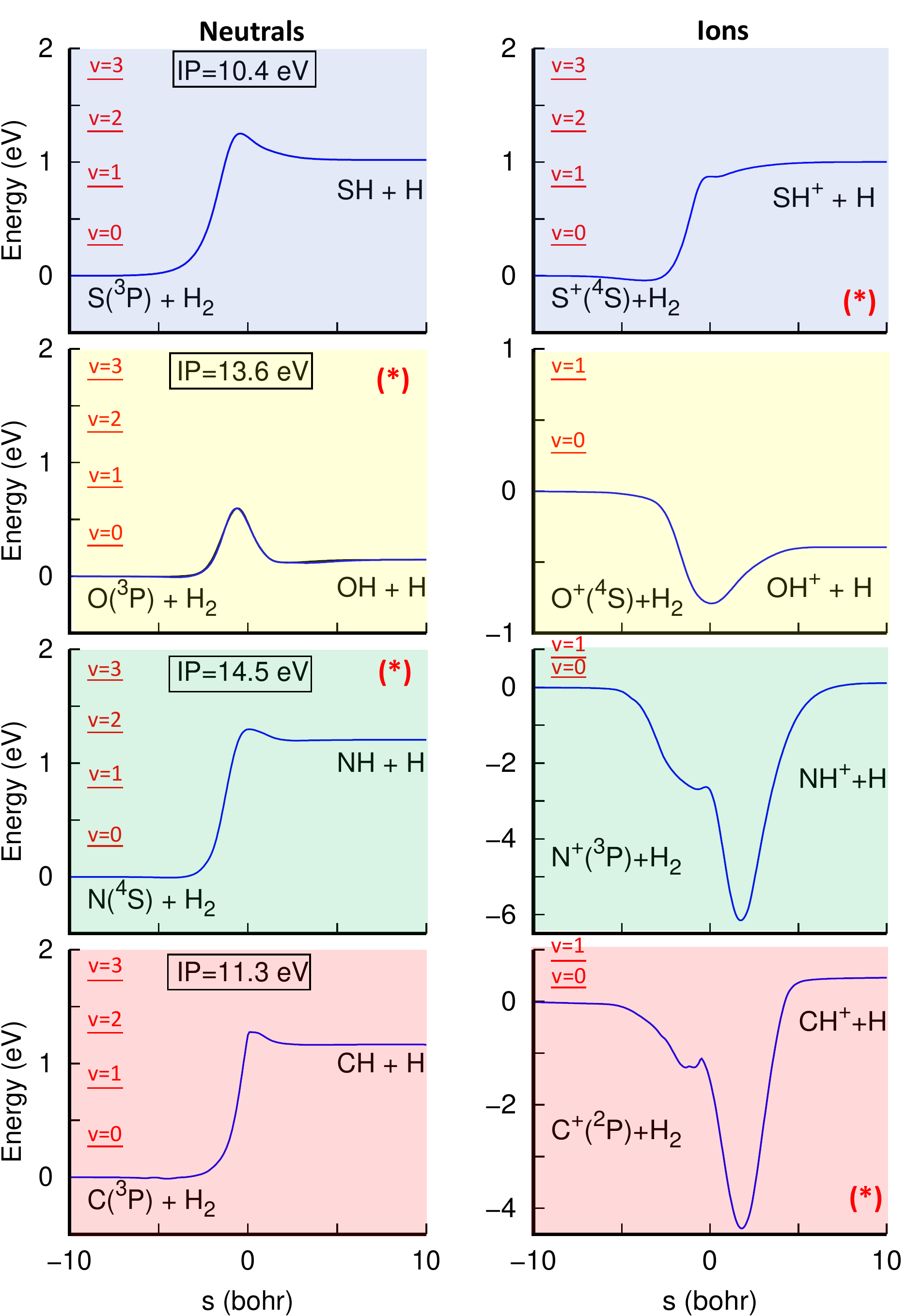}
\caption{Minimun energy paths of key reactions in the ISM involving
H$_2$ and an abundant neutral (left column) and cation element (right column).
In this plot, only the reaction of O$^+$ with \mbox{H$_2$} is exoergic for all $v$.
We define the reaction coordinate as \mbox{$s$\,=\,$r_{\rm H_2}$\,$-$\,$r_{\rm XH}$},
minimizing the energy in the remaining internal degrees of freedom but for a collinear
\mbox{H$-$H$-$X} geometry (except for C$^+$ and N$^+$, in which the angle also varies to
show the depth of the well).  
Each panel shows the energy associated with the lowest-energy vibrationally excited
levels of \mbox{H$_2$\,($v$)} (marked in red). Panels with an asterisk
indicate the dominant ionization state of the element in diffuse neutral clouds
and at PDR edges. 
}
\label{fig:reaction_paths}
\end{figure}

The formation rate of hydride XH through \mbox{reaction~(\ref{reaction-general})} depends on the abundance of element X, the 
 population of H$_2$ in different vibrational levels, and the particular characteristics and dynamics of \mbox{reaction~(\ref{reaction-general})}.
Figure~\ref{fig:reaction_paths} shows
the minimum energy paths  for  reactions of H$_2$ with X.
To make these plots, we took the analytical PESs from different studies:
  \mbox{S($^3P$)\,+\,H$_2$} from \cite{Maiti-etal:04}; 
  \mbox{S$^+$($^4S$)\,+\,H$_2$} from \cite{Zanchet19}; 
  \mbox{O($^3P$)\,+\,H$_2$} from \cite{Zanchet-Menendez-etal:19}; 
  \mbox{O$^+$($^4S$)\,+\,H$_2$} from \cite{Martinez-etal:04};
  \mbox{N($^4S$)\,+\,H$_2$} from \cite{Poveda-Varandas:05};
  \mbox{N$^+$($^3P$)\,+\,H$_2$} from \cite{Nyman-Wilhelmsson:92};
  \mbox{C($^3$P)\,+\,H$_2$} from \cite{Harding93}; and
\mbox{C$^+$($^2P$)\,+\,H$_2$} from \cite{Stoecklin-Halvick:05}.

In general, the endoergicity of reaction~(\ref{reaction-general}) depends on the relative stability of the species  XH with respect to H$_2$ (i.e., their dissociation energy). Reactions involving cations X$^+$ present a diverse behavior: exothermic for O$^+$, endothermic for C$^+$ and S$^+$, and nearly thermoneutral for N$^+$
(although \cite{Zymak-etal:13} report a low endothermicity of 17\,meV that is still under debate).  Some of these reactions present
a  deep insertion well (C$^+$ and N$^+$) and others present a shallow well (O$^+$). However,  none of them present an energy barrier, except for the one attributted to the endothermicity. The lack of energy
 barriers is ultimately  attributed to the presence of
dangling orbitals in the atomic cations. These orbitals allow for the formation of new chemical bonds before  H$_2$ bond dissociation.

Reactions involving neutral atoms, however, display a different and more homogeneous behavior. For all cases where X is in the ground electronic state, which is by far the most
common situation in  molecular clouds, reaction~(\ref{reaction-general}) is
endoergic and possesses an energy barrier due to the lack of  free dangling orbitals of the reactant  $^4S$ and $^3P$  neutral atoms. The absence of dangling orbitals requires,
 first, the fragmentation of the H$_2$ molecule and, second, the formation of the hydride bond. This gives rise to a reaction barrier\footnote{The situation is different
 if one considers the first excited electronic
 states of the following reactants: N($^2D$), O($^1D$), or S($^1D$). In these cases
 a deep insertion well appears, but there is no reaction barrier. Inside molecular clouds
 these elements are largely present in the ground  state.}.

The presence of a well and/or  a barrier along the reaction path introduces an important difference in the reaction mechanisms and, hence, in the reaction rate coefficients. For example, the presence of deep wells between the reactants leads to the formation of long-lived molecular complexes (e.g., CH$_{2}^{+}$) in which there is an important energy transfer among all internal degrees of freedom. 

 During the approach of the two reactants in a reaction with no wells,
 one can consider that there is no flow of energy,  from vibrational or rotational, to the translation degrees of freedom. If there is no barrier,
  the reaction threshold opens as soon as the total energy is above the zero-point energy of the products, thus overcoming
  the endothermicity. However, the presence of a barrier posses constraints depending on how the energy
  is distributed among the degrees of freedom and on the particular location of the barrier.
  According to  Polany's rules 
  for barriers at the entrance channel ({early barrier}), the translational energy between reactants must 
  be above the height of the barrier to overpass it
  \citep{Polanyi-Wong:69,Mok-Polanyi:69}. However,
  when the barrier is located in the products channel ({late barrier}),
  then the vibrational energy has to be higher than the reaction 
  path\footnote{These are qualitative arguments that depend on the involved masses
  and 
  not all can be extrapolated to polyatomic systems.
  For that purpose, \citet{Guo-Jiang:14} developed a semi-quantitative model,
  the ``sudden vector projection model,'' which should be considered
  as approximate since the problem is generally more complex.}.

\section{Ab initio study of reaction {N($^4S$)\,+\,H$_2$($v$)\,$\rightarrow$\,NH\,+\,H}}
\label{sec:N-reaction}

In this section we describe our study of \mbox{reaction~(\ref{reac-NH})}
and  we determine the state-specific rate coefficients up to $v$\,=\,12.
We also compare  the rates of analogous reactions having a different heavy element atom as reactant.

Figure~\ref{fig:reaction_paths} shows that  \mbox{reaction~(\ref{reac-NH})} 
is very endoergic when \mbox{H$_2$} in the ground vibrational state ($v$\,=\,0). 
However, the reaction becomes exoergic
for \mbox{H$_2$\,($v$\,$>$\,3)}. Here we compute the state-specific reaction rates from accurate quantum dynamics calculations employing a wave packet method
for H$_2$($v$\,$\leq$\,7).  For  H$_2$($v$\,$\geq$\,7), quantum calculations become computationally demanding and we employ a quasi-classical trajectory
method (QCT), which is a classical method that uses quantum initial conditions. 
These kinds of reaction dynamics studies require an accurate PES. Here we use
the PES determined by  \cite{Poveda-Varandas:05} from very accurate
\mbox{ab initio} calculations of the ground adiabatic $^4A'$ electronic state.

\subsection{Quantum and quasi-classical  calculations}\label{sub-rates}

\mbox{Following} our previous studies, we used   
the quantum wave packet method implemented in the \mbox{MADWAVE3}
package \citep{MADWAVE3:21,Gomez-Carrasco-Roncero:06,Zanchet-etal:09b}.
We used reactant Jacobi coordinates
 and determined the state-to-state reaction probabilities for each calculated angular momentum \citep{Gomez-Carrasco-Roncero:06}.
 Table\ref{tab:wvp-parameters} lists the parameters used in the calculations.
We performed these  calculations for total angular momentum ($J$ ) and parity
($p$) under the inversion of coordinates, extracting the total reaction
 probability, $P^{Jp}(E)$. The total integral cross section for reaction
 \mbox{N\,+\,H$_2$($v$, $j$=0)\,$\rightarrow$\,NH + H} 
 was then obtained from the following usual partial wave
 expression:
 \begin{eqnarray}
   \sigma_{v}(E) ={ \pi\over (2j+1) k^2_{vj}}\sum_{J,p} (2J+1) P^{Jp}(E),
\end{eqnarray}
where $j$ is the initial rotational state of H$_2$ (here we adopt $j$\,=\,0), $E$ is the translational
energy, and \mbox{$k_{vj}=\sqrt{ 2\mu E\over\hbar^2}$, and $\mu= m_N 2 m_H/(m_N+2m_H)$} is the reduced mass of the
\mbox{N\,+\,H$_2$} system. 

 \begin{table}[t]
 \caption{\label{tab:wvp-parameters}
   Parameters of our quantum calculations using \mbox{MADWAVE3} in reactant Jacobi coordinates:
   $r_{min} \leq r\leq r_{max}$ is the H$_2$ internuclear distance,
   $R_{min} \leq R\leq R_{max}$ is the distance between the H$_2$ center-of-mass and the nitrogen
   atom, and $0 \leq \gamma \leq \pi/2$ is the angle between ${\vec r}$ and ${\vec R}$ vectors.
   The initial wave packet
   is described in $R$ by a Gaussian centered at $R=R_0$ at an translational energy of
   $E=E_0$, and width $\Delta E $. The total reaction probability was obtained by analyzing the total flux
   at $r=r_\infty$
}
 \begin{center}
 \begin{tabular}{cc}
 \hline \hline
 $r_{min}$, $r_{max}=$  0.1, 24 \AA & $N_r$=320 \\
 $r_{abs}$=  11 \AA & \\
$R_{min}$, $R_{max}=$   0.001, 24\AA & $N_R$=512  \\
 $R_{abs}$=  11 \AA  &\\
$N_\gamma$ = 140 & in $[0,\pi/2]$  \\
$R_0$  = 10 \AA & $E_0,\Delta E$= 1,0.4 eV\\
$r_\infty$ = 4 \AA &   \\
 \hline
 \end{tabular}
 \end{center}
 \end{table}

We calculated the reaction probabilities for 
\mbox{$J$\,=\,0,\,10, ... ,\,80}.
We employed the $J$-shifting-interpolation and extrapolation method
\citep{Aslan-etal:12} to obtain the reaction probabilities of the intermediate $J$ values
by evaluating the total integral
reactive cross section using the partial wave expansion up to $J$=80.
\mbox{Finally}, we determined the state-specific reaction rate coefficients
 by numerical integration of the cross section with a Boltzmann distribution for the translational
 energy \citep{Zanchet13b,Gomez-Carrasco-etal:14}. That is to say, 
  \begin{equation}
  k_v(T) =  \left[\frac{8}{\pi \,\mu \,(k_{\rm B}\,T)^3} \right]^{1/2} Q_e(T) \,
  \int_{0}^{\infty} E\,\sigma_v(E)\,e^{-E/k_{\rm B}T}\,dE,
  \label{eqn:rates}
\end{equation}
 where $Q_e(T)$ is the electronic partition function. 
 
\begin{figure*}[t]
\centering   
\includegraphics[scale=0.64, angle=0]{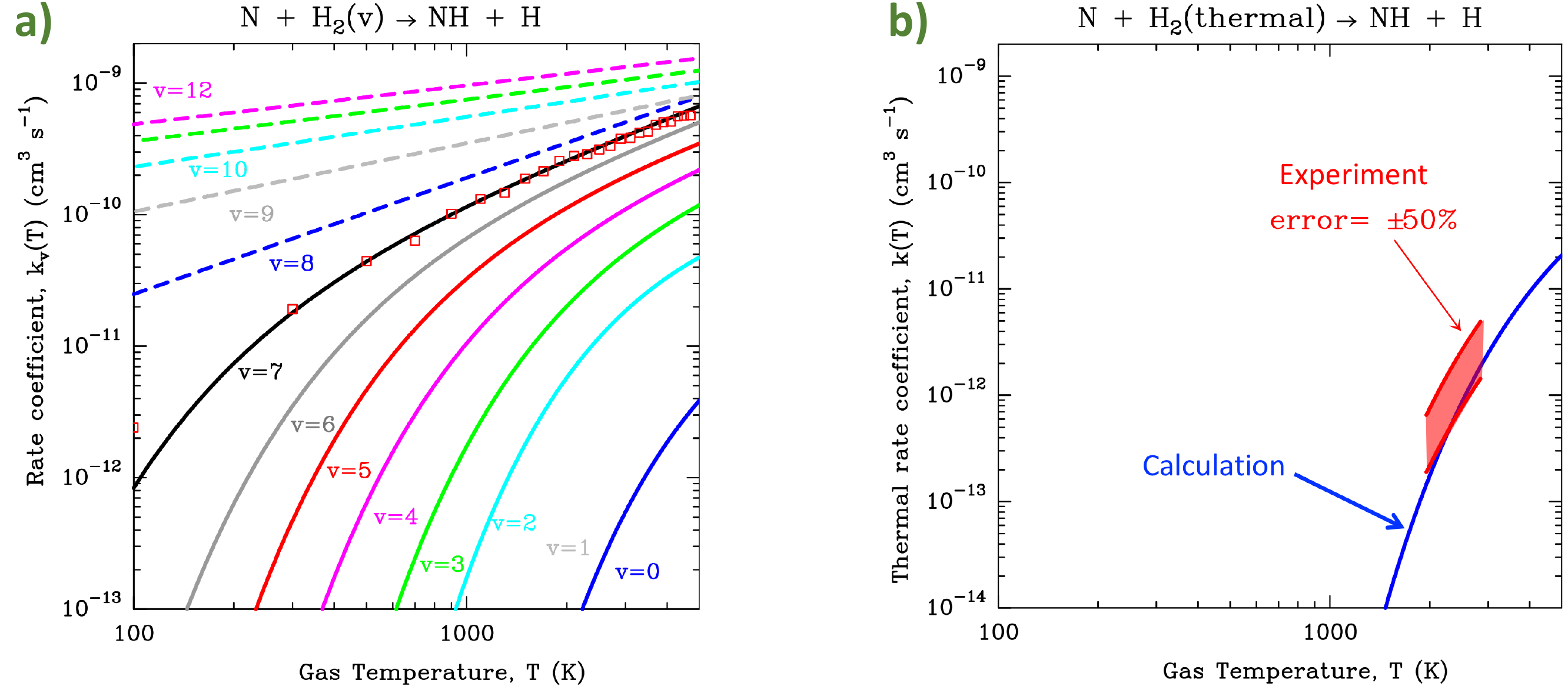}
\caption{Calculated rate coefficients of  reaction \mbox{N($^4$S)\,+\,H$_2$($^1$$\Sigma_{g}^{+}$,$v$)\,$\rightarrow$\,NH($^3$$\Sigma^{-}$)\,+\,H}.
\textit{Left panel}: H$_2$ vibrational-state-specific  rates from wave packet calculations
(continuous curves) and QCT calculations (dashed curves). Red squares show how QCT calculations
reproduce the quantum calculations for $v$\,=\,7.
\textit{Right panel}: Thermal rate coefficient (blue curve) calculated from thermal averages of the
state-specific rate coefficients. The red shaded area represents the experimental values determined by
\citet{Davidson1990} allowing for an experimental error of $\pm$50\,\%. }
\label{fig:reaction_rates}
\end{figure*}

For H$_2$($v\geq7$), we performed  quasi-classical calculations using the  \mbox{MDwQT} code \citep{Sanz-Sanz-etal:15,Zanchet-etal:16,Ocana-etal:17}.
We sampled the initial conditions with the usual
 Monte Carlo method \citep{Karplus-etal:65}.
 For each vibrational state of  H$_2$, the initial internuclear distance and velocity distributions were
 obtained with the adiabatic switching method \citep{Grozdanov-Solovev:82,Qu-Bowman:16,Nagy-Lendvay:17}. 
 The initial rotational state of H$_2$ was set to zero. The distance between N and H$_2$ center of mass
 was initially set to 50\,bohr, and we stopped the trajectories when the internuclear distance was longer than 60\,bohr.
 The initial impact partameter was sampled between 0 an 15\,bohr, according to a quadratic distribution.
 For each translational temperature, we ran between 2$\times$10$^4$ and 10$^5$ trajectories for each $v$ and $T$.
 The lower temperatures (\mbox{$T$\,$\simeq$\,100\,K}) and vibrational states  $v$\,=\,7 and 8 require the largest number of trajectories.
 We determined the QCT state-specific rate coefficients as follows:
  \begin{equation}
  k_v(T) =  \sqrt{\frac{8 k_B T}{\pi \mu}} \pi b_{max}^2(T)P_r(T)
  \label{eqn:k_T},
\end{equation}
where $b_{max}(T)$ is the maximum impact parameter and $P_r(T)$
is the reaction probability at a constant temperature.

\subsection{State-specific reaction rates and discussion}

Figure~\ref{fig:reaction_rates}$a$ shows the calculated state-specific 
rate coefficients from \mbox{$v$\,=\,0} to 12. Continuous curves show the
quantum calculations and dashed lines show the QCT calculations. We specifically compared the  quantum (curves) and classical (red squares) rate coefficients for \mbox{H$_2$($v$\,=\,7)}. They show very good agreement except at \mbox{100 K}, at which the reaction probability is very low and thus the statistical error is large. This supports the validity of the QCT rates obtained for \mbox{$v>$\,7}.
The right panel of Fig.~\ref{fig:reaction_rates} shows the resulting thermal rate
coefficient computed as the Boltzmann average of the individual state-specific coefficients:
  \begin{equation}
  k_{\rm th}(T) =  \frac{\sum_{v=0}^{12}\,k_v(T) \,e^{-E_v/k_{\rm B}T} }
                        {\sum_{v=0}^{12}\,e^{-E_v/k_{\rm B}T}}.  
  \label{eqn:thermal_rate}
\end{equation}
Most astrochemical models use $k_{\rm th}(T)$. This is the rate coefficient commonly provided
by theorists and experimentalists. At low temperatures ($T$\,$\ll$\,$E_v\,({\rm H_2})/k_{B}$), in most \mbox{interstellar} applications, the thermal rate coefficient is roughly that of H$_2$ in the ground vibrational state: 
\mbox{$k_{\rm th}(T)$\,$\simeq$\,$k_{v=0}(T)$} \citep{Agundez10}.
\mbox{Figure~\ref{fig:reaction_rates}$b$} also shows the available experimental
rate coefficient of \mbox{reaction~(\ref{reac-NH})} in the temperature range $T$\,=\,1950-2850\,K \citep{Davidson1990}.
For  \mbox{$T\,>$\,2300\,K}, our calculated thermal rate coefficient lies within the relatively large
experimental error of the laboratory measurement (red shaded area). For \mbox{$T\,<$\,2300\,K}, our calculated rate is slightly below the experimental error. We attribute these slight differences to the fact that we neglect the rotational levels in the Boltzmann average. Still, we are confident that the overall good agreement validates the results of our calculations.

\begin{table}[t]
\centering
\caption{H$_2$ vibrational energies ($E_v$) and Arrhenius-like fit parameters,
 $k_v(T) = \alpha \,(T/300\,{\rm K})^{\beta} e^{-\gamma/T}$, of the state-specific rate coefficients  calculated  for
 reaction \mbox{N($^4S$)\,+\,H$_2$($v$)\,$\rightarrow$\,NH\,+\,H}. }
\begin{tabular}{c  c r c r}
  \hline\hline
  $v$ & $E_v$ (eV) & $\alpha$ (cm$^3$\,s$^{-1}$) & $\beta$ & $\gamma$ (K)  \\
  \hline
           0 &  0.270    &   0.721\,$\times$\,10$^{-10}$ &   0.000     &  14629.000 \\
           1 &  0.784    &   1.631\,$\times$\,10$^{-10}$ &   0.000     &  10161.600 \\
           2 &  1.270    &   1.922\,$\times$\,10$^{-10}$ &   0.000     &   7005.160 \\
           3 &  1.727    &   0.501\,$\times$\,10$^{-10}$ &  0.596      &   4078.190 \\
           4 &  2.156    &   0.410\,$\times$\,10$^{-10}$ &  0.758      &   2273.080 \\
           5 &  2.557    &   0.537\,$\times$\,10$^{-10}$ &  0.766      &   1420.000 \\
           6 &  2.931    &   0.534\,$\times$\,10$^{-10}$ &  0.856      &    817.174 \\
           7 &  3.275    &   0.501\,$\times$\,10$^{-10}$ &  0.943      &    305.879 \\
           8 &  3.599    &  6.581\,$\times$\,10$^{-11}$  &  0.885      &      0.000 \\
           9 &  3.884    &  1.873\,$\times$\,10$^{-10}$  & 0.520       &      0.000 \\
          10 &  4.134    &  3.516\,$\times$\,10$^{-10}$  & 0.380       &  0.000 \\
          11 &  4.349    &  5.132\,$\times$\,10$^{-10}$ & 0.317        &       0.000  \\
            12 &  4.523  &  6.757\,$\times$\,10$^{-10}$ & 0.295        & 0.000   \\\hline
              & thermal  & 5.000\,$\times$\,10$^{-10}$ & 0.000  &  15900.000 \\  
\hline
 \end{tabular}
\label{tab:rates-parameters} 
\end{table}

Table~\ref{tab:rates-parameters} lists the obtained state-specific reaction rate coefficients
fitted by the usual Arrhenius-like form 
\mbox{$k_v(T)\,=\,\alpha\,(T/300\,{\rm K})^{\beta}\,{\rm exp}(-\gamma/T)$}.
We note that these rate coefficients increase with temperature and 
with the initial  H$_2$ vibrational level $v$.
For H$_2$($v$\,$\geq$\,3), the reaction becomes exoergic, but the reaction rate still behaves as if it has an energy threshold due to the location of the energy barrier in the PES.

\begin{figure}[t]
\centering   
\includegraphics[scale=0.59, angle=0]{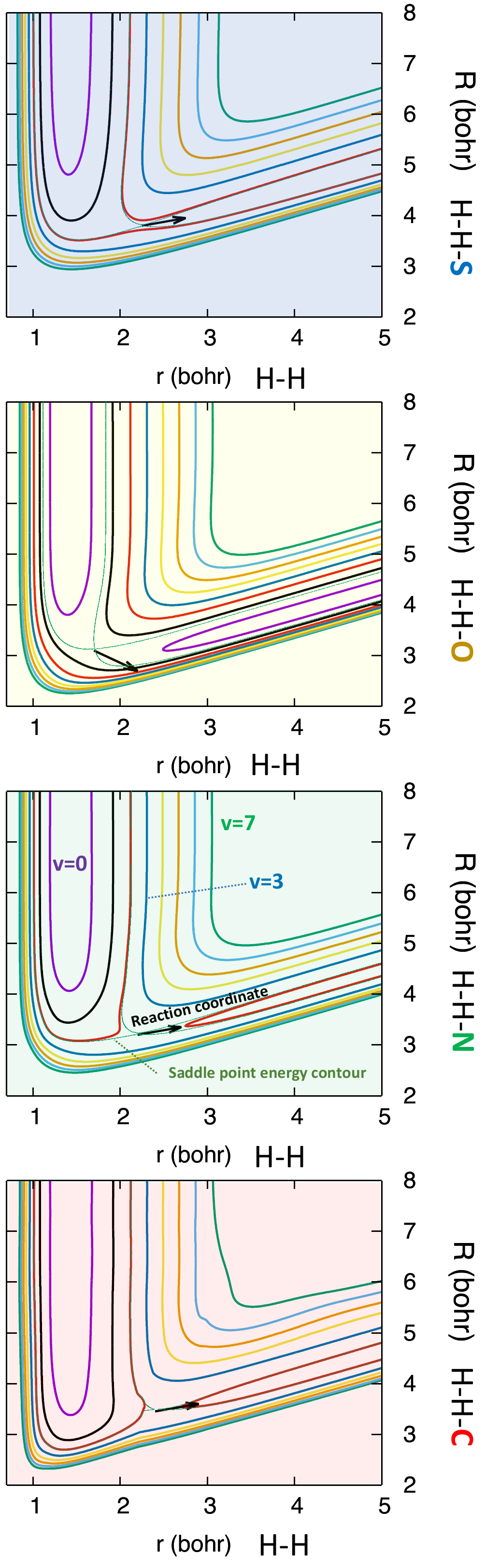}
\caption{Contour plot of the potential energy surfaces
describing reactions 
\mbox{X\,+\,H$_2$\,$\rightarrow$\,XH\,+\,H} (with X\,=\,C, N, O, and S)
as a function of $r$, the H$_2$ internuclear distance, and $R$, 
the distance between the N and H$_2$ center-of-mass
in a colinear H-H-X configuration. 
Contours correspond to the vibrational energies of H$_2$ from 
$v$\,=\,0 to 7. The thin green line shows
the saddle point energy contour.
The thick black arrow is a  graphical representation of the reaction coordinate at the top of the barrier.}
\label{fig:pes2d}
\end{figure}

Figure~\ref{fig:pes2d} (second panel from the bottom) shows the PES of a  colinear H-H-N configuration as a function of 
the Jacobi distances $R$, the distance between the N and the H$_2$ center-of-mass  (whose derivative
is proportional to the initial translational velocity), and also as a function of
 $r$, the H$_2$ internuclear distance, thus associated
with the vibrational energy.
 In this plot, the top of the barrier is along the thick black arrow,
 which is a
graphical approximation of the reactive coordinate at the saddle point.

This saddle point corresponds
to a long $r$ distance, that is to say a late barrier. This means that in order to overcome the energy barrier,
 some vibrational energy needs to be given to the $r$ coordinate. The forth blue contour plot corresponds to the energy of H$_2$($v$\,=\,3)
and it is open, thus connecting reactants and products. Interestingly, one would expect that for H$_2$($v$\,=\,3),
the reaction becomes exoergic without any threshold. However, this is not the case because  the
 direction of the reaction coordinate at the top of the barrier is 
only approximately
parallel to the arrow (Figure~\ref{fig:pes2d}). With an arrow parallel to the 
\mbox{$x$ axis}, the slope to overpass
the barrier would have been along the $r$ coordinate, and the reaction with H$_2$($v$\,=\,3)  would be roughly \mbox{exoergic}.
However, the arrow has a  non-negligle contribution along the $R$ coordinate. Hence, 
some energy must be given to the system along this $R$ coordinate  initially 
\mbox{(i.e., translational energy)}
to overcome the barrier. This late energy barrier implies that the reaction rate coefficients are very small
at low temperatures but  increase with $T$ and with the vibrational state even for very \mbox{high-$v$} H$_2$ levels.

\begin{figure}[t]
\centering   
\includegraphics[scale=0.55, angle=0]{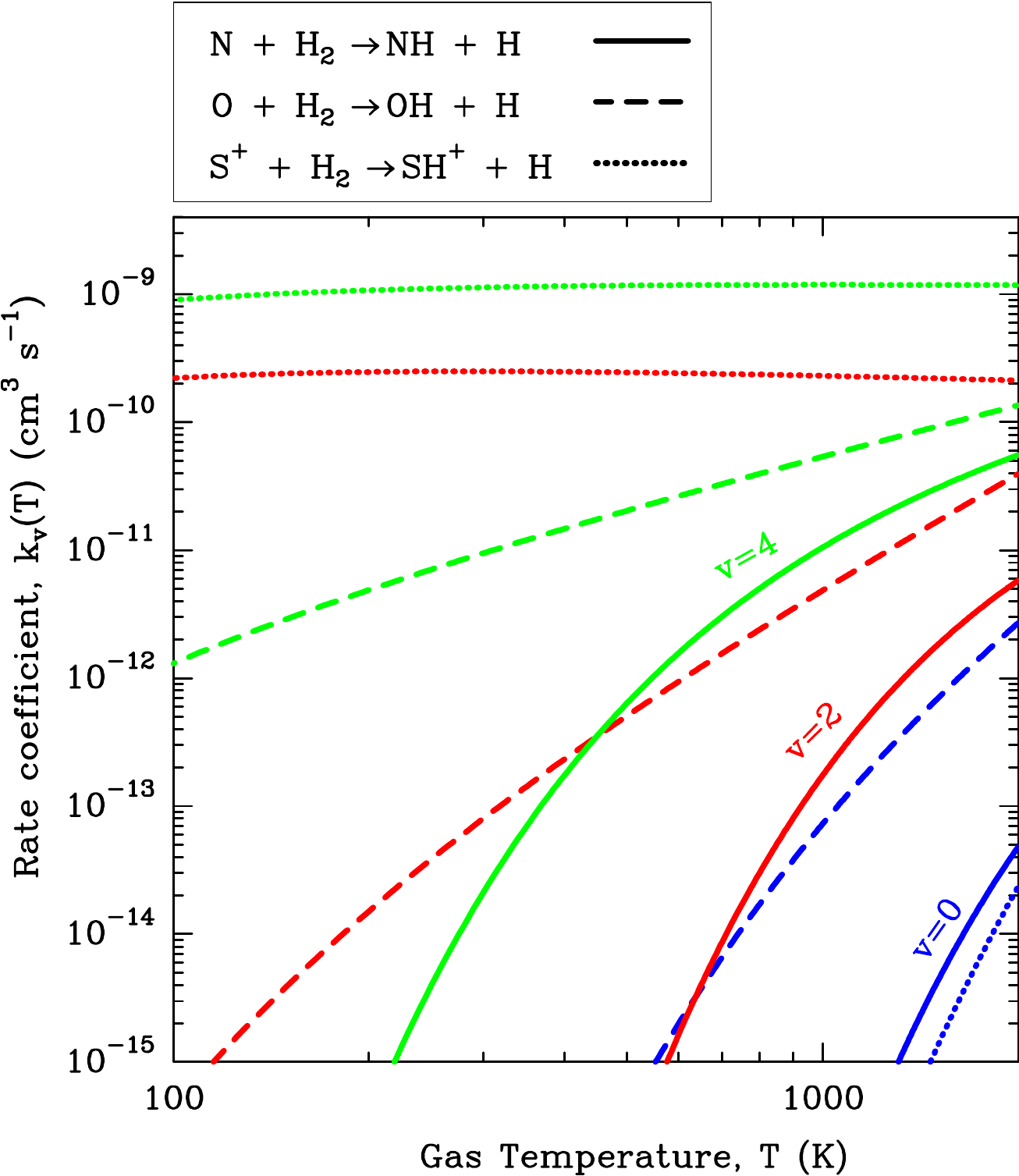}
\caption{Comparison of H$_2$ ($v$\,=\,0, 2, and 4) state-specific rate
coefficients  of reactions \mbox{N($^4S$)\,+\,H$_2(v)$\,$\rightarrow$\,NH\,+\,H} (continuous curves, this work),
\mbox{O($^3P$)\,+\,H$_2(v)$\,$\rightarrow$\,OH\,+\,H} \citep[dashed curves,][]{Veselinova21},
and \mbox{S$^+$($^4S$)\,+\,H$_2(v)$\,$\rightarrow$\,SH$^+$\,+\,H} 
 \citep[dotted curves,][]{Zanchet19}, all calculated from quantum methods. }
\label{fig:compa_rates}
\end{figure}

\subsubsection{Comparison with analogous  reactions} 

Figure~\ref{fig:pes2d} summarizes the qualitatively similar PESs
of  reactions between neutral atomic N, S, C, and O  and vibrationally excited H$_2$.
These reactions show a similar behavior, all being endothermic with a late barrier and no well.
The moderate increase in the state-specific rate coefficients $k_v(T)$ with $v$  and
temperature is produced by the presence of these barriers.
On the other hand, endothermic hydrogen abstraction reactions involving
N$^+$, S$^+$, and C$^+$, thus lacking energy barriers,
behave differently. For example, reaction \mbox{S$^+$\,+\,H$_2$($v$)$\rightarrow$\,SH$^+$\,+\,H}  (Fig.~\ref{fig:reaction_paths}) becomes exoergic for \mbox{$v$\,$\geq$\,2} and
its associated rate coefficients do not show any threshold for \mbox{$v$\,$\geq$\,2}
\citep{Zanchet19}. That is,
they do not vary with temperature much.
Figure~\ref{fig:compa_rates} explicitly compares  the \mbox{$v$\,=\,0, 2, and 4} state-specific rate coefficients of reactions between H$_2$($v$) and N($^4S$), S$^+$($^4S$), and O($^3P$) \citep[the latter is from][]{Veselinova21}. 
\mbox{As expected, the} rate coefficients of reaction \mbox{S$^+$\,+\,H$_2$($v$)}
 for
\mbox{$v$\,$\geq$\,2} are nearly independent of temperature. However, the rate coefficients of reaction \mbox{O\,+\,H$_2$($v$)}
 still increase with $T$. A relevant difference between hydrogen abstraction reactions with O and N is the lower height of the energy barrier for reaction \mbox{O($^3P$)\,+\,H$_2$($v$)}. This translates into higher rate
coefficients for the lower--$v$ H$_2$ vibrational states, and thus
a stronger dependence on the production of interstellar OH, compared to NH,  with gas temperature 
\citep{Veselinova21}.
As we demonstrate in the next section, the formation of abundant NH specifically  requires
\mbox{FUV-pumped} H$_2$ and the subsequent population of highly excited $v$ levels. These processes take place at the irradiated surface of PDRs where  elements
with \mbox{IP\,$>$\,IP(H)}, such as O and N, are in predominantly  neutral form. 
On the other hand,  reactions of neutral atomic carbon and sulfur with vibrationally excited H$_2$
are less important because neutral atomic C and S only become abundant deeper inside the PDR,  where the FUV radiation field is more attenuated and the fraction of
H$_2$ in vibrationally excited states drastically  decreases. 
For comparison purposes, in \mbox{Appendix~\ref{Sect:Appendix}} we calculate QCT state-specific rate coefficients of the analogous reaction {\mbox{S($^3P$)\,+\,H$_2$($v$\,=\,0-12)\,$\rightarrow$\,SH\,+\,H$_2$}
(see \mbox{Table~\ref{tab:rates-parameters-S}}). Despite the similar behavior of 
the rate coefficients of reactions  \mbox{N($^4S$)\,+\,H$_2$($v$)}
and \mbox{S($^3P$)\,+\,H$_2$($v$)}, the impact of the latter ones on the 
formation of SH radicals is very minor.
The difference arises from the fact the S$^+$ ions, and not S atoms, are the dominant sulfur reservoir in the PDR layers where vibrationally excited H$_2$ is abundant
(see Fig.~\ref{fig:mods-S}).

\section{Effects of the state-specific rates on the NH abundance in FUV-irradiated gas}\label{sec:PDR_models}

Dense PDRs contain enhanced densities of \mbox{FUV-pumped} H$_2$.  Infrared
observations of the Orion Bar show the presence of H$_2$ ro-vibrational emission 
lines from excited levels up to \mbox{$v$\,=\,12} \mbox{\citep[e.g.,][]{Kaplan21}}.
The relative populations of these highly excited levels depart from thermal (purely collisional) excitation. 
 Here we study the reactivity 
of highly vibrationally excited H$_2$ with N  atoms in \mbox{illumination}  conditions
appropriate to two iconic PDRs, the Orion Bar
 and the rim of the Horsehead nebula.
Their impinging FUV radiation fields are 
\mbox{$G_0$\,$\simeq$\,2$\times$10$^4$} \citep{Marconi98}  and 
\mbox{$G_0$\,$\simeq$\,100} \citep{Abergel03}, respectively.

\begin{table}[t]
\caption{Main parameters used in the PDR models of the Orion Bar.\label{table:PDR-mods}} 
\centering
\begin{tabular}{ccc@{\vrule height 8pt depth 5pt width 0pt}}
\hline\hline
Model parameter                                 &     Orion Bar                          &     Horsehead        \\ 
\hline
FUV radiation field, $G_0$                                      &     2$\times$10$^4$ Habing\,$^{(a)}$   &  100 Habing\,$^{(b)}$           \\
Thermal pressure $P_{\rm th}/k_{\rm B}$         &  2$\times$10$^8$\,cm$^{-3}$K\,$^{(c)}$ &  4$\times$10$^6$\,cm$^{-3}$K\,$^{(d)}$\\\hline
 $n_{\rm H}$\,=\,$n$(H)\,+\,2$n$(H$_2$)         & \multicolumn{2}{c}{$n_{\rm H}$\,=\,$P_{\rm th}\,/\,k_{\rm B}T_{\rm k}$}    \\
Cosmic Ray $\zeta_{\rm CR}$                     & \multicolumn{2}{c}{10$^{-16}$\,H$_2$\,s$^{-1}$}                              \\
$R_{\rm V}$\,=\,$A_{\rm V}$/$E_{\rm B-V}$       & \multicolumn{2}{c}{5.5\,$^{(e)}$}                                 \\
Abundance O\,/\,H                               & \multicolumn{2}{c}{3.2$\times$10$^{-4}$\,$^{(f)}$}                           \\
Abundance C\,/\,H                                                       & \multicolumn{2}{c}{1.4$\times$10$^{-4}$\,$^{(g)}$}                     \\
Abundance N\,/\,H                                                               & \multicolumn{2}{c}{7.5$\times$10$^{-5}$\,$^{(h)}$}                                        \\
Abundance S\,/\,H                                                           & \multicolumn{2}{c}{1.4$\times$10$^{-5}$\,$^{(i)}$}                    \\
\hline                                    
\end{tabular}
\tablefoot{$^a$\cite{Marconi98}. $^b$\cite{Abergel03}. $^c$\citet{Joblin18}. $^d$\citet{Habart06}.  $^e$\citet{Cardelli89}. $^f$\cite{Meyer98}. $^g$\cite{Sofia04}.
 $^h$\cite{Meyer97}. $^i$\cite{Goico21}.}\label{Table:PDR_parameters}
\end{table}

To model the nitrogen chemistry in these two PDRs, we used  the Meudon PDR 
code  \citep{LePetit06}. 
 We  implemented the state-specific rates of \mbox{reaction~(\ref{reac-NH})} calculated in \mbox{Sect.~\ref{sec:N-reaction}}  up to \mbox{H$_2$\,($v=12$)}.
In addition, we updated the rate coefficient of the destruction reaction
\mbox{NH\,+\,H\,$\rightarrow$\,N\,+\,H$_2$} (relevant in PDRs due to the enhanced abundance
of hydrogen atoms). We adopted the rate coefficient computed by \cite{Han10}, which is
consistent with laboratory experiments of \cite{Adam05}.
Our models also include specific reactions of $o-$H$_2$ and  $p-$H$_2$ 
with N$^+$ ions, for which we computed the H$_2$ ortho-to-para (OTP) ratio  at each cloud depth.
We adopted the rate coefficients fitted from low-temperature ion trap experiments of 
 \cite{Zymak-etal:13}. The reaction \mbox{N$^+$\,+\,H$_2$} 
is thought to initiate\footnote{The rate coefficient of reaction \mbox{N$^+$($^3P_J$)\,+\,H$_2$($j$)\,$\rightarrow$\,NH$^+$\,+\,H} depends on the
 H$_2$($j$) rotational level population  (i.e., on the OTP ratio). At low  temperatures,
reactions with \mbox{$o$-H$_2$} are faster \citep[e.g.,][]{Marquette_88,Zymak-etal:13},
but \mbox{$o$-H$_2$} is also less abundant.
In our models, we ignore the different reactivities of N$^+$ in its three $^3P_J$ fine structure levels (separated by 
\mbox{$\Delta E_{\rm FS}$\,/\,$k_B$\,=\,70\,K} and
188\,K, respectively). 
\cite{Zymak-etal:13} derived the rate coefficients considering that only the three
lower spin-orbit states contribute to the reaction and that 
the highest excited N$^+$ fine-structure state $^3P_2$ is not reactive (adiabatic behavior).
Such an approximation changes  the determination of the reaction
rate coefficients for \mbox{$o$-H$_2$} and \mbox{$p$-H$_2$}, especially at low temperatures. Recent quantum calculations  including  transitions among all spin-orbit states find that
the reactivity of the state  $^3P_2$  is not zero \citep{Gomez-Carrasco_22}. 
This implies a slight reduction in the rate coefficients  for  $o$-H$_2$, especially at low temperatures. This agrees with new experimental results of
\cite{Fanghaenel_18}, which take the contribution of the  $^3P_2$ state into account.} 
 the nitrogen
 chemistry in cold gas shielded from FUV radiation \citep[e.g.,][]{LeBourlot_91,Dislaire_12}.
 The PDR model performs a detailed   
treatment of the  H$_2$ \mbox{FUV-pumping} and vibrational excitation  as well as of the penetration of FUV radiation into the cloud \mbox{\citep{Goicoechea07}}. Since we are mainly interested in the most irradiated outer layers of the PDR, 
where H$_2$ molecules are effectively pumped by FUV photons,
 our models  only include gas-phase chemistry \citep[except for  H$_2$  formation;][]{Bron14}. Following our previous studies,
we use a constant thermal pressure ($P_{\rm th}/k_{\rm B}$\,=\,$n_{\rm H}$\,$T$). 
Table~\ref{Table:PDR_parameters} summarizes the main  parameters
and gas-phase abundances adopted in our models.

\subsection{Strongly FUV-irradiated gas: The Orion Bar PDR case}\label{subsec:PDR_Bar}

Figure~\ref{fig:collage_bar} dissects the  physical structure of the Orion Bar
model as a function of cloud depth  (along the illumination direction and in magnitudes of visual extinction, $A_V$).
Figure~\ref{fig:collage_bar}$a$ shows the decreasing gas temperature  and 
increasing H$_2$ density gradient from the PDR edge to the more \mbox{FUV-shielded} cloud interior. It is important to note that 
\mbox{H$_2$ molecules} are efficiently photodissociated at the irradiated PDR surface  \mbox{($A_V$\,$\lesssim$\,1\,mag)}. In these hot gas layers, 
the abundance of H atoms is higher than that of H$_2$.
However, a significant fraction of the existing H$_2$ is in highly excited vibrational states
(shown in Fig.~\ref{fig:collage_bar}$b$). In particular, the fractional abundance of H$_2$ molecules in
$v>7$ states ($f_7$) with respect to those in the ground $v=0$  reaches 
\mbox{$f_7$\,$\simeq$\,3\%}. On the other hand, beyond \mbox{$A_V$\,$\gtrsim$\,2\,mag,} most hydrogen is in a molecular form, but
$f_7$ becomes negligible and the fractional abundance of H$_2$ in vibrationally excited states \mbox{($n$(H$_2$\,$v$\,$\geq$\,1)\,/\,$n_{\rm H}$)} sharply
declines below $\sim$10$^{-8}$. 
Therefore, we expect that any enhanced formation of XH hydrides through 
reactions of X  with highly vibrationally excited H$_2$  will take place in
these \mbox{$A_V$\,$<$\,2\,mag}  surface layers.

Figure~\ref{fig:collage_bar}$c$ shows \mbox{$k_v(T) \times f(v)$}, the contribution of each  H$_2$ vibrational state  to the \mbox{N\,+\,H$_2$($v$)} rate coefficient, as a function
of cloud depth. Here $f(v)$
 is the fractional population of H$_2$ in the vibrational level $v$.
This plot demonstrates that the formation of NH radicals through 
\mbox{reaction~(\ref{reac-NH})} is dominated by
H$_2$ in highly vibrationally excited states \mbox{($v$\,$>$\,7)}.
To visualize the impact of the state-specific rate coefficients of
\mbox{reaction~(\ref{reac-NH})}, Fig.~\ref{fig:collage_bar}$d$ shows the resulting 
 NH  abundance profile (\mbox{$x$(NH)\,=\,$n$(NH)\,/\,$n_{\rm H}$}; blue curves, right axis) and the total  \mbox{N\,+\,H$_2$} rate coefficient  (black curves, left axis). Continuous  curves refer to a PDR model that includes  state-specific rate coefficients, whereas the dashed curves show a model that uses  the thermal rate. Any difference between the continuous and
dashed curves is produced by the nonthermal populations of the highly vibrationally
excited levels of H$_2$.

Remarkable differences appear even at the PDR surface (\mbox{$A_V$\,$<$\,0.1\,mag}), where the fraction of H$_2$ molecules in highly vibrationally states  is large. 
 These still semi-atomic \mbox{($x$(H)\,/\,$x$(H$_2$)\,$\gg$\,1)} hot gas   layers show $x$(NH) abundances   about two orders of magnitude higher than the predictions of the model that uses the thermal rate coefficient.
 The enhancement of the NH abundance  is very large even in hot gas
($T_{\rm k}$\,$\simeq$\,2000\,K in this model) because NH formation
is driven by the highly excited vibrationally H$_2$ states that 
cannot be thermally populated.

 The predicted NH abundance peak, \mbox{$x$(NH)\,$\simeq$\,4\,$\times$\,10$^{-9}$}, is located at 
 \mbox{$A_V$\,$\simeq$\,0.8\,mag}. 
 Beyond   \mbox{$A_V$\,$\gtrsim$\,2\,mag}, the amount of H$_2$ in vibrationally excited states sharply declines and both models  predict the same NH abundance. 
 The inclusion of state-specific rates for \mbox{reaction~(\ref{reac-NH})}
increases the total NH column density (integrated from $A_V$\,=\,0 to 10\,mag) by at factor of $\sim$25, from \mbox{$N$(NH)\,=\,2.1$\times$10$^{11}$\,cm$^{-2}$} to \mbox{\,5.5$\times$10$^{12}$\,cm$^{-2}$}.  According to these models, most of the NH column density
is located close to the PDR surface, between $A_V$\,$\simeq$\,0.5 and 2\,mag.

\begin{figure}[t] 
\centering   
\includegraphics[width=0.855\linewidth]{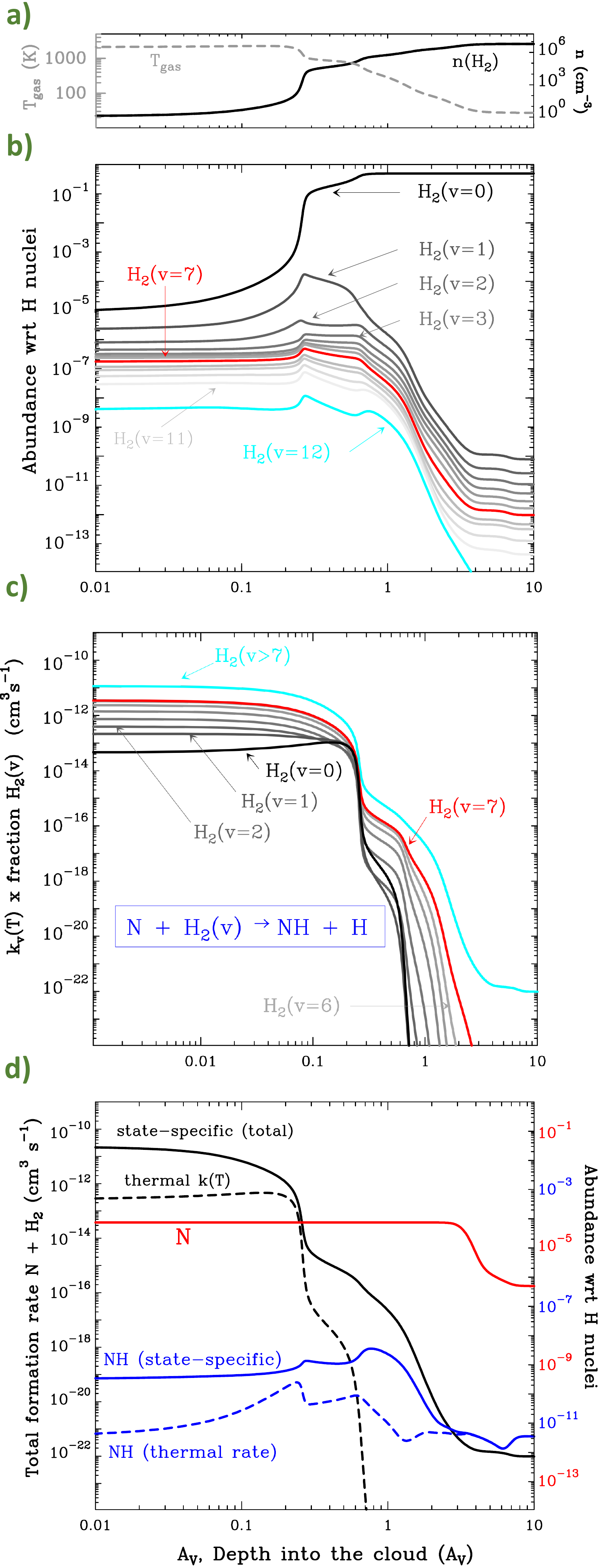}

\caption{Physical structure and quantities of the Orion Bar  model as a function of visual extinction.
\mbox{\textit{a)} Gas temperature} and H$_2$ density.
\textit{b)} Fractional abundances of each H$_2$ vibrational level.
\textit{c)} Contribution of each individual H$_2$ vibrational level to
the \mbox{N\,+\,H$_2$($v$)} reaction rate coefficient and denoted as
\mbox{$k_v$($T$)\,$\times$\,$f(v)$}, with  
\mbox{$f(v)$\,=\,$n$(H$_2$,\,$v$)\,/\,$n$(H$_2$)}.
\mbox{\textit{d)} Total} formation rate coefficient (black curves) and  N and NH abundance profiles.}
\label{fig:collage_bar}
\end{figure}


Figure~\ref{fig:HH_Bar_HH_models}$a$ shows the predicted $x$(N), 
$x$(N$_2$), $x$(NH), $x$(N$^+$), $x$(NH$^+$), and
\mbox{$x$(H$_2$\,($v$\,=\,7))}  abundance profiles, 
and the fraction of H$_2$ in $v$\,$>$\,7 states ($f_7$).
Owing to the high ionization potential of neutral N atoms, \mbox{IP(N)\,$>$\,IP(H)}, neutral N atoms are the major nitrogen reservoir at the PDR surface and up to \mbox{$A_V$\,$\simeq$\,4\,mag}.
In the most irradiated PDR layers, even if the abundance of H$_2$
with respect to H nuclei is still low,
NH readily forms by reactions of N with highly vibrationally excited H$_2$ (which is the main 
destruction pathway of N in these layers).  Destruction of NH is initially driven by 
reactive collisions with abundant H atoms \mbox{(NH\,+\,H\,$\rightarrow$\,N\,+\,H$_2$)}.
At the NH abundance peak, the density of H atoms decreases 
(upper panel of Fig.~\ref{fig:HH_Bar_HH_models}$a$), but the FUV radiation field
is still high. Hence, photodissociation dominates NH destruction.
This happens because  NH  does not easily react with H$_2$. Indeed, 
reaction \mbox{NH\,+\,H$_2$\,$\rightarrow$\,NH$_2$\,+\,H} is endothermic and possesses 
a high activation barrier, \mbox{$E_{\rm A}$/\,$k_{\rm B}$\,$\simeq$\,5700\,K} \mbox{\citep{Linder95}}.
Although NH$_2$ is not detected in the Bar, the role of the above reaction might be more relevant for \mbox{FUV-pumped} H$_2$. However, state-specific
reaction rates have not been computed yet.

\begin{figure*}[t]
\centering   
\includegraphics[scale=0.465, angle=0]{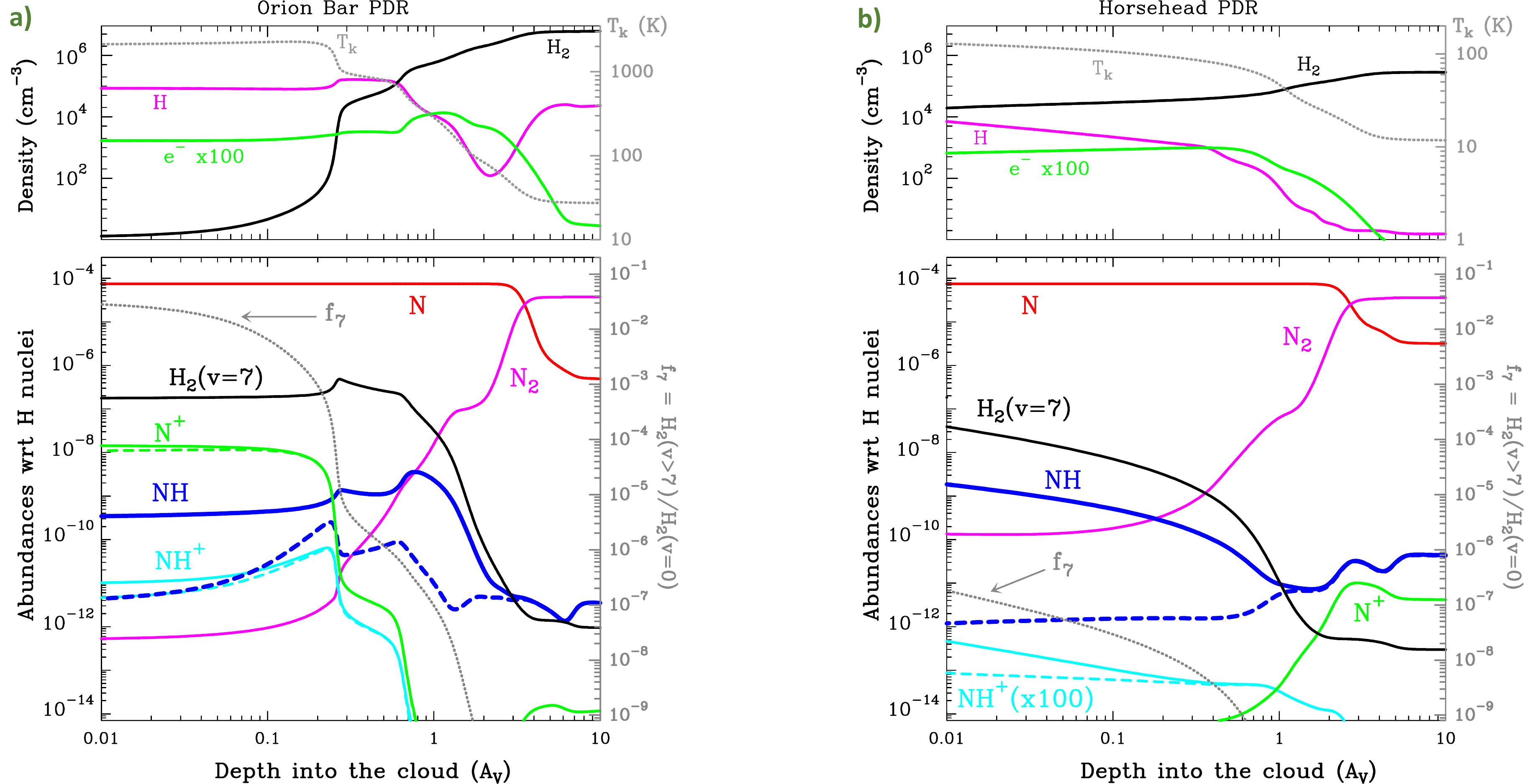}
\caption{Isobaric models of the Orion Bar  (\mbox{$G_0$\,$\simeq$2$\times$10$^4$},  
\mbox{$P_{\rm th}/k_{\rm B}$\,=\,2$\times$10$^8$\,cm$^{-3}$\,K}) and of the Horsehead
PDR 
(\mbox{$G_0$\,$\simeq$\,100}, \mbox{$P_{\rm th}/k_{\rm B}$\,=\,4$\times$10$^6$\,cm$^{-3}$\,K}).
\mbox{\textit{Upper panels}:} Density and gas temperature structure
as a function of visual extinction from
the PDR surface, $A_V$.
\textit{Lower panels}: Abundance profiles with respect to H nuclei. The gray dotted curve shows $f_7$, the fraction of H$_2$($v>7$) with respect to the ground (right axis gray scale).
Solid curves refer to a model using state-specific reaction rates
for \mbox{reaction~(\ref{reac-NH}),} whereas dashed curves refer to a model using the thermal rate.}
\label{fig:HH_Bar_HH_models}
\end{figure*}

Beyond the NH abundance peak, at $A_V$\,$\gtrsim$\,2\,mag, most H$_2$ is in the
lowest-energy vibrational states and NH formation  through 
\mbox{reaction~(\ref{reac-NH})} becomes negligible (see the evolution of the total formation rate in Fig.~\ref{fig:collage_bar}$d$). 
At this point,
other chemical reactions -- notably those initiated by the hydrogen abstraction reactions
\mbox{N$^+$\,$\rightarrow$\,NH$^+$\,$\rightarrow$\,NH$_{2}^{+}$\,$\rightarrow$\,NH$_{3}^{+}$}
and finishing with the dissociative recombination of NH$_{3}^{+}$ and NH$_{2}^{+}$ ions --
dominate the production of NH \mbox{\citep[e.g.,][]{Boger05,Dislaire_12}}. 
In the Orion Bar model, the number of N$^+$ ions is controlled by the photodissociation of N-bearing 
ions (such as NO$^+$ and NH$^+$) at the PDR edge, and by the cosmic-ray ionization rate
at large cloud depths (through direct ionizations of N atoms and through reactions of N$_2$ with He$^+$). 
\mbox{Figure~\ref{fig:Reaction_evol}} shows the contribution (in~percent)
of reactions \mbox{N\,+\,H$_2$($v$)} and \mbox{N$^+$\,+\,$o$/$p$-H$_2$} to the total
formation rate of NH and NH$^+$, respectively. At the PDR edge, $A_V$\,$\lesssim$\,2\,mag,  the production of NH is almost entirely dominated by reaction \mbox{N\,+\,H$_2$($v$)}.
 The reaction  \mbox{N$^+$\,+\,H$_2$} and  NH photoionization contribute to the production of  NH$^+$.
However,  due to the lower abundance of N$^+$ ions compared to that
of neutral N  atoms, the \mbox{NH\,/\,NH$^+$} column density ratio is significantly above one.
Deeper inside the cloud, as FUV radiation  decreases,  N$_2$ becomes the main nitrogen reservoir (at \mbox{$A_V$\,$>$\,4\,mag} in this model).  Here, the nitrogen chemistry is initiated by reactions of N$^+$ with H$_2$, which is 
sensitive  to the H$_2$ OTP ratio (red curve in \mbox{Fig.~\ref{fig:Reaction_evol})}.
At larger $A_V$, grain  chemistry associated with 
the formation and desorption of ammonia ice \mbox{\citep[][]{Knacke82,Wagenblast93}} may alter the dominant formation and destruction pathways of NH.

\subsection{Mildly FUV-illuminated gas: The Horsehead PDR case}\label{subsec:PDR_HH}

Figure~\ref{fig:HH_Bar_HH_models}$b$ shows the abundance profiles predicted by
our model of the Horsehead PDR (less intense FUV field and lower gas density). 
 Only at the  PDR surface (A$_V$\,$<$\,0.5\,mag)
is the abundance of H$_2$ (with respect to H nuclei) in highly vibrationally excited
states  large enough, 
with \mbox{$x$(H$_2$,\,$v$=7)\,$\simeq$\,5\,$\times$\,10$^{-10}$} at 
\mbox{$A_V$\,$\simeq$0.5\,mag} (versus \mbox{$\simeq$\,3\,$\times$\,10$^{-7}$} in the  Bar). In addition, the fraction of H$_2$ in highly
vibrationally excited levels is lower:
 $f_7$\,$\simeq$\,2\,$\times$\,10$^{-9}$  at  \mbox{$A_V$\,$\simeq$0.5\,mag} versus $f_7$\,$\simeq$\,10$^{-6}$ in the Orion Bar. 
The higher  $x$(NH) abundance  at the very surface of the Horsehead PDR
compared to that in the Bar  is a consequence of the lower density of H atoms 
and lower flux of FUV photons (i.e., lower NH destruction rate).
 
With respect to the nitrogen chemistry, the use of state-specific rates for \mbox{reaction~(\ref{reac-NH})} does increase the production
of NH by two orders of magnitude  at the Horsehead  surface.
However, the enhancement factor quickly decreases deeper inside the PDR because the abundance of H$_2$\,($v$\,$\gg$) diminishes with $A_V$  being steeper than in a strongly irradiated dense PDR, such as the Bar. 
The total NH column density predicted by the Horsehead model, integrated from $A_V$\,=\,0 to 10\,mag, is \mbox{$N$(NH)\,=\,1.2$\times$10$^{12}$\,cm$^{-2}$}
\mbox{(i.e., 4.5 times lower  than in the Orion Bar)}.
Compared to the Bar, the edge of Horsehead  has lower
abundances of N$^+$ ions, which are solely formed by cosmic-ray ionization
of  N atoms and not by the photoionization of N-bearing molecular ions
(their abundances are very low). On the other hand, the abundance of N$^+$
 at large $A_V$   is higher in the Horsehead PDR  than in the Orion Bar (where the gas is warmer, $n$(H$_2$) is higher, and  thus N$^+$ ions are more easily destroyed).

\begin{figure}[t]
\centering   
\includegraphics[scale=0.45, angle=0]{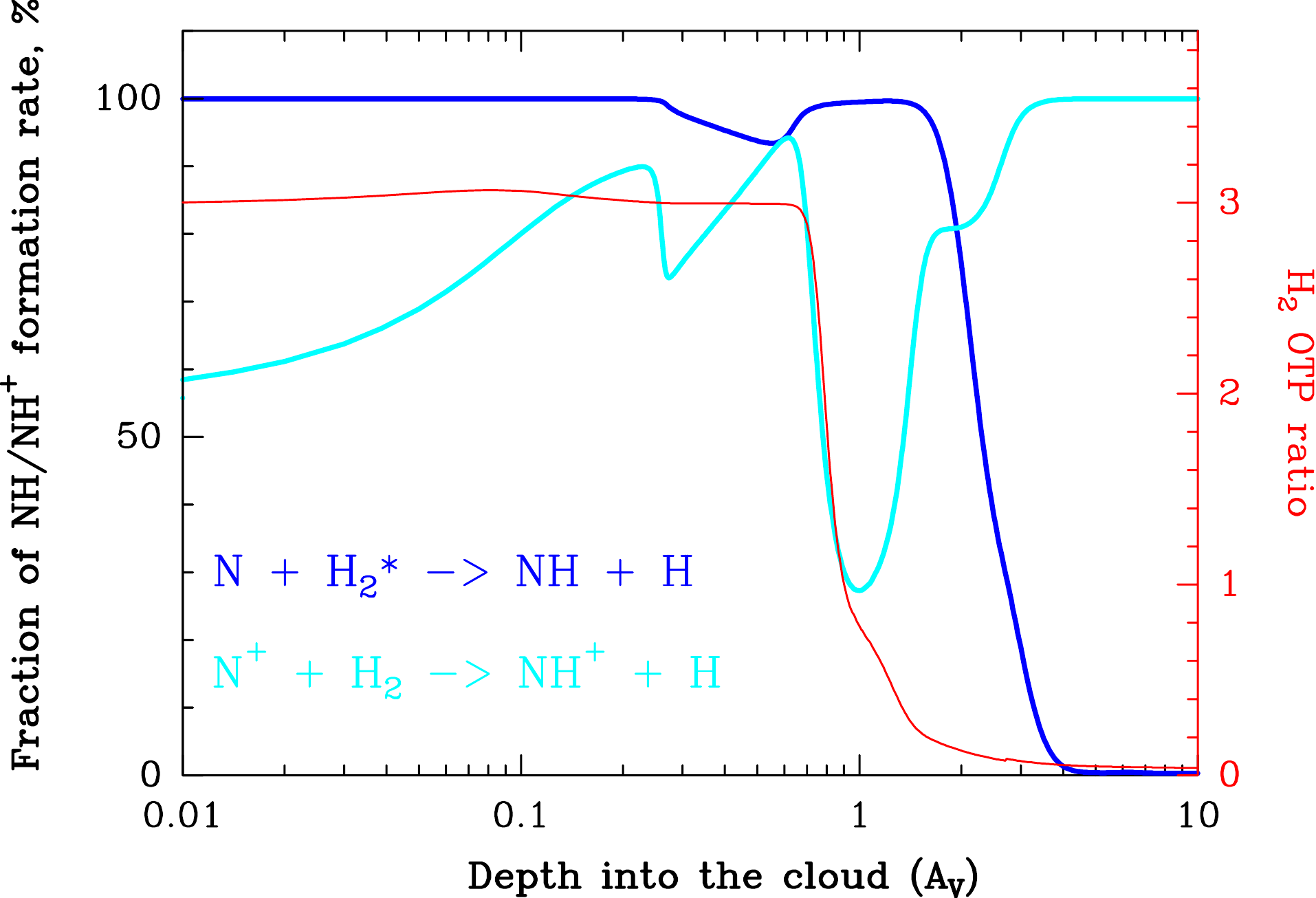}
\caption{H$_2$  OTP ratio (red) and contribution (in~percent)  of  reactions \mbox{N\,+\,H$_2$($v$)\,$\rightarrow$\,NH\,+\,H} and
\mbox{N$^+$\,+\,$o$/$p$--H$_2$\,$\rightarrow$\,NH$^+$\,+\,H}
to the total formation rate of NH (blue) and NH$^+$ (cyan) in the Orion Bar model.}
\label{fig:Reaction_evol}
\end{figure}

\section{Detectability of submillimeter NH emission lines}\label{sec:Detectability}

In this section we study the excitation and detectability of  NH rotational \mbox{lines} in dense PDRs. We also 
explore existing high-spectral resolution submillimeter  observations of the Orion Bar and Horsehead 
and search for NH  emission lines.

The electronic ground state of the NH radical is $^3\Sigma^-$. Hence, rotational levels
with $N>0$ show a triplet fine structure arrangement \citep[for an energy diagram, see][]{Klaus97}. In addition, the \mbox{$I_{\rm H}$\,=\,0.5} and  \mbox{$I_{\rm N}$\,=\,1} \mbox{nuclear} spins lead to a hyperfine splitting of the fine structure levels. 
Being a light molecule, the lowest-energy rotational lines (\mbox{$N$\,=\,$1-0$}) 
appear in the high-frequency submillimeter domain, at 946\,GHz ($N_J=1_0-0_1$),  974\,GHz ($N_J=1_2-0_1$), and 1000\,GHz ($N_J=1_1-0_1$). 
High resolution heterodyne observations allow one
to detect and spectrally resolve some of these NH hyperfine structure (HFS) lines
\citep{Persson10,Persson12}. 
Unfortunately,  telluric absorption precludes the observation of theses lines from ground-based telescopes.

\subsection{Subthermal NH emission in  PDRs}\label{subsec:excitation}

Inelastic collisions  \mbox{populate} the rotationally excited  molecular levels
in dense molecular gas.
However, NH is a hydride molecule with a large rotational constant and high
spontaneous radiative decay  rates, \mbox{$A_{ij}$($N$\,=\,$1-0$)\,$\simeq$} several 10$^{-3}$\,s$^{-1}$. This is about 10$^5$ times larger than those of the widely observed \mbox{CO $J$\,=\,$1-0$} line. 
This means that
the gas density at which the inelastic collision de-excitation rate coefficients ($\gamma_{ij}$ in cm$^3$\,s$^{-1}$) equal the spontaneous radiative emission rate
-- the so-called critical density
 \mbox{$n_{\rm cr}$\,=\,$A_{ij}$\,/\,$\gamma_{ij}$($T_{\rm k})$}  of a given rotational transition -- is much
higher than the gas density of the medium. For  NH, 
\mbox{$n_{\rm cr}$($N$\,=\,$1-0$)\,$\simeq$\,10$^9$\,cm$^{-3}$}. This implies
that even in dense molecular clouds (\mbox{$n_{\rm H}$\,$>$\,10$^4$\,cm$^{-3}$}), 
NH rotational lines will be  weakly (subthermally) excited\footnote{To estimate the NH--H$_2$ and NH--H
inelastic collision rate coefficients, we simply scaled the available fine-structure-resolved \mbox{NH--He}  rate coefficients
(computed by \citet{Tobola11} up to level \mbox{$N$\,=\,8} and \mbox{$T_{\rm k}$\,=\,350\,K}) 
by the square root of the reduced mass ratio. We also extrapolated them to higher
temperatures.}. In other words, \mbox{$n_{\rm cr}$\,$\ll$\,$n_{\rm H}$} results in $T_{\rm rot}\ll T_{\rm k}$.
Together with the intricate  NH HFS spectrum (the rotational partition function of NH at 150\,K is $\sim$15 times larger than that of CH$^+$), this means that, unless exceptionally  abundant, interstellar NH {emission} lines will be faint.

Figure~\ref{fig:MTC_models} shows the rotational temperature of the NH \mbox{$N_J=1_2-0_1$} fine-structure
line (974\,GHz) as a function of gas temperature  for different H$_2$ densities.
These curves are \mbox{``single-slab''} model results obtained from detailed  
nonlocal \mbox{thermodynamic} equilibrium (LTE) excitation calculations$^5$ using a  Monte Carlo code 
\citep[][and references therein]{Goicoechea22}. They refer to optically thin NH line emission.
The inspection of \mbox{Fig.~\ref{fig:MTC_models}} shows that one can expect 
  \mbox{$T_{\rm rot}$(974\,GHz)\,$\simeq$\,10--15\,K}
in the NH emitting layers of the Bar (a hot and dense PDR; \mbox{Fig.~\ref{fig:HH_Bar_HH_models}a})
and \mbox{$T_{\rm rot}$(974\,GHz)\,$\simeq$\,4-6\,K} in the cooler lower density Horsehead (\mbox{Fig.~\ref{fig:HH_Bar_HH_models}b}).


\subsection{3$\sigma$ detection  of NH in the Orion Bar PDR}\label{subsec:detection}

The Orion Bar and \mbox{Horsehead}  were observed at the frequencies of NH \mbox{($N$\,=\,$1-0$)} lines  with the
HIFI receiver  \citep{deGra10} on board Herschel. These observations are part of the HEXOS \mbox{(PI. E. A. Bergin)} and 
WADI \mbox{(PI. V. Ossenkopf)} guaranteed time key programs\footnote{The NH $N_J=1_0-0_1$ lines at 946\,GHz 
appear in HIFI \mbox{band 3b}  (observed in the Bar \mbox{[ObsID 1342216380]}). The NH \mbox{$N_J=1_2-0_1$} lines at 974\,GHz  appear in HIFI band 4a (observed in the Bar \mbox{[ObsID 1342218628]} and in the Horsehead PDR \mbox{[ObsID 1342218215]}).
The NH \mbox{$N_J=1_1-0_1$} lines at 1000\,GHz appear in HIFI band 4a (observed in the Orion Bar 
\mbox{[ObsID 1342218628]}).}. The Orion Bar data belong to a fully calibrated line survey
\mbox{\citep[][]{Nagy17}}. The Horsehead data belong to a deeper search of hydride molecules. These HIFI spectra  around the \mbox{NH ($N$\,=\,$1-0$)} lines have not been discussed in the literature.
In principle, the spectral resolution (1.1\,MHz or $\sim$0.3\,km\,s$^{-1}$ at these frequencies) is high enough
to resolve several HFS lines\footnote{We took the hyperfine levels and spectroscopic information tabulated in 
the Cologne Database for Molecular Spectroscopy \citep[CDMS;][]{Endres16}. These data contain
various experimental spectra \citep{Klaus97,Flores-Mijangos04,Lewen04}.} of the three fine structure transitions  $N_J=1_0-0_1$ (946\,GHz),  $N_J=1_2-0_1$ (974\,GHz), and $N_J=1_1-0_1$ (1000\,GHz).

\begin{figure}[t]
\centering   
\includegraphics[scale=0.67, angle=0]{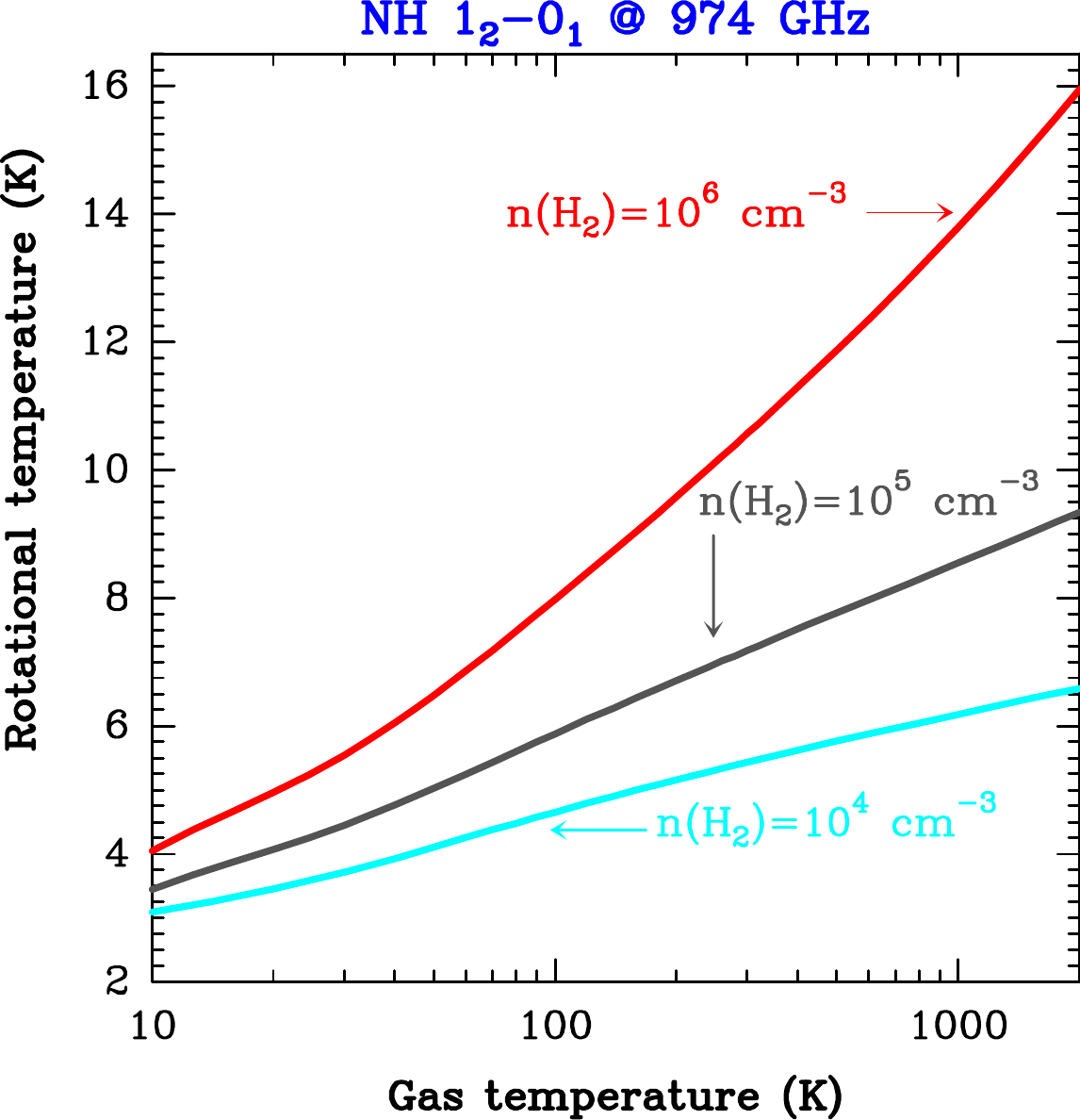}
\caption{Rotational temperature of the NH $N_J$\,=\,1$_2$-0$_1$ fine-structure transition 
versus gas kinetic temperature, for different gas
densities, obtained from non-LTE excitation and optically thin emission models.}
\label{fig:MTC_models}
\end{figure}

\begin{figure}[t]
\centering   
\includegraphics[scale=0.9, angle=0]{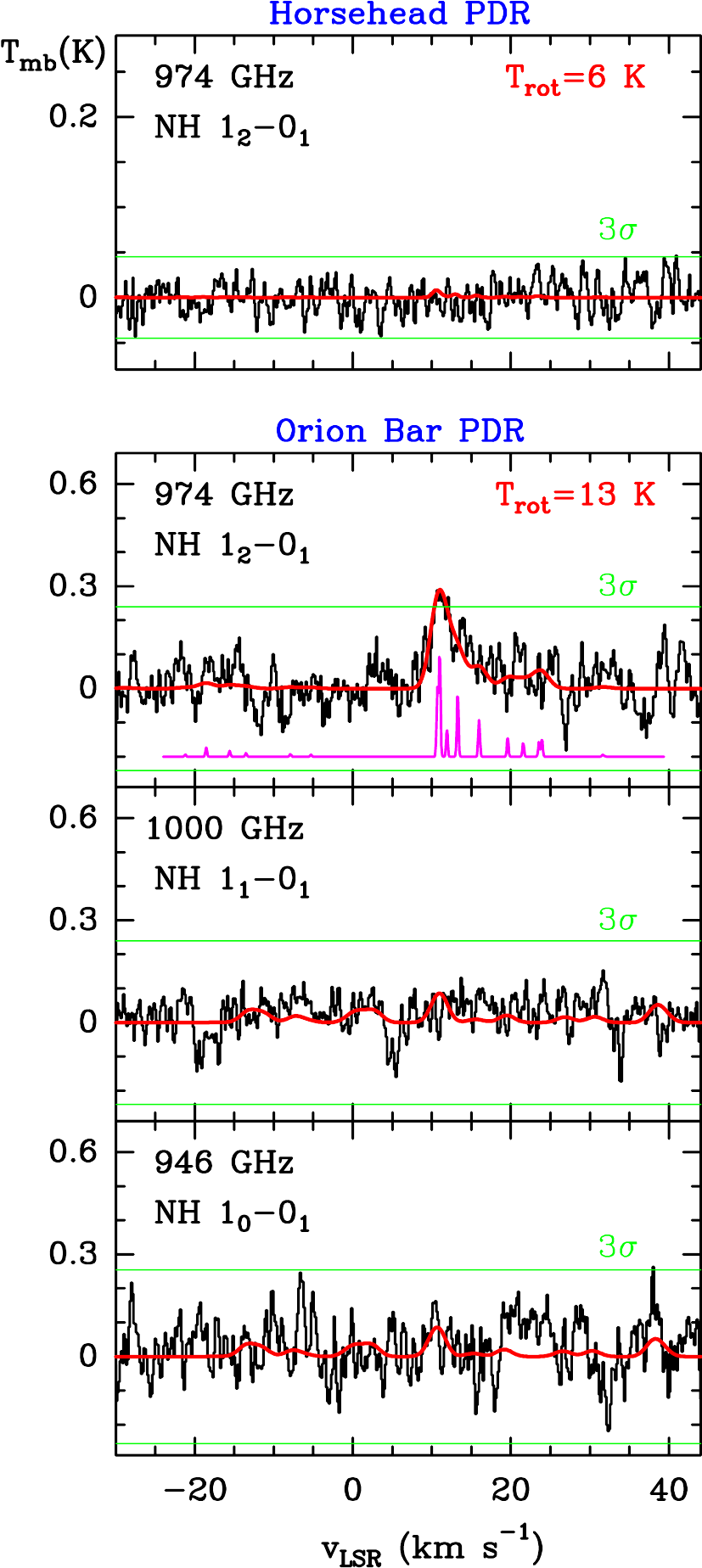}
\caption{Herschel/HIFI observations of the Horsehead and Orion Bar PDRs.
The green horizontal lines display the $\pm$3$\sigma$ rms  levels.
  Red curves are single-slab 
models for \mbox{$N$(NH)\,=\,1.3\,$\times$\,10$^{13}$\,cm$^{-3}$}.
These data show a  3$\sigma$ detection (at the line peak) at 974.479\,GHz.
The pink spectrum shows the relative strengths of the individual HFS components
for optically thin line emission.
The  model of the Bar is also consistent
with the nondetection of the fainter NH lines at $\sim$946 and $\sim$1000\,GHz.}
\label{fig:NH_HIFI}
\end{figure}

\textit{The Orion Bar}:  Fig.~\ref{fig:NH_HIFI} shows  HIFI observations of the Bar. The root mean square (rms) noise of these
spectra is 80-90\,mK per velocity channel. The 974\,GHz spectra show an emission feature
above the 3$\sigma$ noise level that  matches the velocity of the molecular emission
in the Orion Bar \citep[v$_{\rm LSR}$\,$\simeq$10.7\,km\,s$^{-1}$; e.g.,][]{Cuadrado15} 
if the feature is at $\sim$974.479\,GHz. This is exactly the frequency of the brightest 
\mbox{$F_1 =5/2-3/2$}, \mbox{$F=7/2-5/2$} (974478.38\,MHz) and
\mbox{$F_1 =3/2-1/2$}, \mbox{$F=5/2-3/2$} (974479.34\,MHz)  HFS components of the
 $N_J=1_2-0_1$  transition\footnote{We searched for other possible molecular carriers
 of this feature. The \mbox{HCN $J$\,=\,11$-$10} line is
 at 974487.20 GHz (at only \mbox{$-$2.5\,km\,s$^{-1}$} of the observed emission).
 However,  the lower excitation and lower frequency lines \mbox{HCN $J$\,=\,10$-$9} and
 \mbox{9$-$8} are not detected  \citep{Nagy17}. Hence, the
 $J$\,=\,11-10 line cannot be responsible for this  feature. }. The gas velocity dispersion in the Orion
 Bar is such that the molecular emission observed by single-dish telescopes shows typical line widths of 
 $\Delta$$v$\,$\simeq$\,3\,km\,s$^{-1}$ \mbox{\citep[e.g.,][]{Cuadrado15,Nagy17}}. This implies that the
 width and shape of the observed feature is a blend of several HFS lines
 (individually shown by the pink spectra in the 974\,GHz panel of Fig.~\ref{fig:NH_HIFI}).
 
We modeled the NH HFS spectrum of the Orion Bar adopting \mbox{$T_{\rm rot}$\,=\,13\,K}  and 
 a nonthermal gas velocity dispersion of \mbox{$\sigma_{\rm nth}$\,=\,1~km\,s$^{-1}$} 
 ($\Delta v_{\rm nth}$\,=\,\,2.355$\sigma_{\rm nth}$). We reproduced the amplitude  of the
 observed feature at 974\,GHz with \mbox{$N$(NH)\,=\,1.3$\times$10$^{13}$\,cm$^{-2}$}.
 The red curve in  Fig.~\ref{fig:NH_HIFI} shows the resulting synthetic NH spectra.
Given the low signal-to-noise ratio\footnote{In terms of the integrated line intensity, the detection significance of the $\sim$974.479\,GHz feature over a 10\,km\,s$^{-1}$ wide window is $\sim$5$\sigma$.} (S/N) of the data, the model is also consistent with the nondetection of the fainter $N_J=1_0-0_1$ (946\,GHz)
and  $N_J=1_1-0_1$ (1000\,GHz) HFS lines. 
The data do not show any features at the frequency (1012\,GHz) of 
the NH$^+$ $^2\Pi_{1/2}$ $N=1-1$ $J=3/2-1/2$ line either.
  
\textit{Horsehead}:  The upper panel of Fig.~\ref{fig:NH_HIFI} shows HIFI 
observations at 974\,GHz. Despite this spectrum having a higher sensitivity (rms
of 15\,mK per channel) than that of the Bar, it does not show any emission features above 3$\sigma$.
We created a synthetic spectrum adopting  \mbox{$N$(NH)\,$=$\,1.3$\times$10$^{13}$\,cm$^{-2}$},   \mbox{$T_{\rm rot}$\,=\,6\,K},
 and
\mbox{$\sigma_{\rm nth}$\,=\,0.4~km\,s$^{-1}$}, consistent with the 
typical line width of molecular
lines in the Horsehead PDR \citep[$\Delta$v\,$\lesssim$\,1\,km\,s$^{-1}$; e.g.,][]{Pety12}.
Because of the lower rotational temperatures, the $N_J=1_2-0_1$ (974\,GHz) emission lines will be much fainter in  the Horsehead PDR,  below the sensitivity achieved in these observations.
Hence, the $N$(NH) adopted in the  radiative transfer model just reflects an upper limit value in the Horsehead PDR.

\section{Discussion}

In order to take into account the predicted gradients of the NH emitting layers in a PDR, here we use the full cloud-depth dependent $n$(H), $n$(H$_2$), $T_k$, and $x$(NH) profiles
of the PDR model calculation \mbox{(Fig.~\ref{fig:HH_Bar_HH_models})} as inputs of a non-LTE  radiative transfer  multislab model of NH rotational lines.
Table~\ref{Table:Column_densities} shows the total NH column density of each PDR model as well as the resulting \mbox{NH\,1$_2$-0$_1$} line intensity (integrating
from $A_V$\,=\,0 to 10\,mag).

As anticipated, the predicted NH emission in the Horsehead PDR is below the sensitivity reached by Herschel observations. For the Orion Bar, only the PDR model including state-specific reaction rate coefficients is consistent with the observed level of NH emission, especially if one considers that the PDR is not completely edge-on and one allows for a small tilt angle of 
$\alpha$\,$\lesssim$\,20$^o$ with respect to a fully \mbox{edge-on} geometry
\citep[][]{Jansen95,Melnick12,Andree17}.
This geometry implies that optically thin lines are limb-brightened, with an intensity
enhancement of sin$^{-1}$$\alpha$ with respect to a \mbox{face-on} PDR.
The  NH emission detected in the Bar is consistent with its formation by gas-phase reactions of N atoms with highly vibrationally excited H$_2$
at the PDR surface and producing \mbox{$N$(NH)\,$\simeq$\,10$^{13}$\,cm$^{-2}$}.
Compared to other PDRs, the higher gas density in the  Bar  contributes to a more efficient collisional excitation of submillimeter NH rotational lines, thus leading to detectable emission lines.  The nondetection of NH$^+$, however, is consistent
with the low column density predicted by PDR models, that is to say $N$(NH$^+$) of several 10$^{10}$\,cm$^{-2}$.

\begin{table}[t]
\caption{NH column density and line intensity predictions from multislab radiative transfer models 
(from $A_V$\,=\,0 to 10\,mag).\label{table:PDR-mods}} 
\centering
\begin{tabular}{c c c @{\vrule height 8pt depth 5pt width 0pt}}
\hline\hline
            Input                                          &    \multicolumn{2}{c}{\hspace{1.4cm}$N$(NH) (cm$^{-2}$)}  \\ 
  PDR model &        Orion Bar                            &  Horsehead              \\
\hline
$k_{\rm th}(T)$   &   $2.1\times10^{11\,(a)}-6.3\times10^{11\,(b)}$ & $7.1\times10^{11\,(a)}$         \\
$k_{v=0-12}$($T$) &   $5.5\times10^{12\,(a)}-1.6\times10^{13\,(b)}$ & $1.2\times10^{12\,(a)}$            \\\hline\\
       Input                                             &     \multicolumn{2}{c}{\hspace{1.4cm} W(NH  at 974\,GHz) (mK\,km\,s$^{-1}$)}\\ 
 PDR model &        Orion Bar                            &  Horsehead    \\\hline
$k_{\rm th}(T)$   &  5$^{(a)}-15^{(b)}$   &    11$^{(a)}$      \\
$k_{v=0-12}$($T$) &  290$^{(a)}-870^{(b)}$  &   17$^{(a)}$          \\\hline
Observations      &  $\sim$\,700       &    $<$\,25   \\

\hline                                    
\end{tabular}
\tablefoot{$(a)$ For a face-on PDR. $(b)$ Edge-on PDR with a tilt angle $\alpha$\,=\,20$^o$  (geometrical
enhancement of a factor of three).}\label{Table:Column_densities}
\end{table}

Our study implies that the detection of NH emission  in PDRs traces strongly \mbox{FUV-irradiated} dense 
gas\footnote{Submillimeter NH lines are also detected in absorption toward the dust continuum  emitted by massive 
\citep{Fuente10,Bruderer10,Benz10} and low-mass  \citep{Hily-Blant10,Bacmann10}
star-forming cores.
These detections refer to lower density (\mbox{$n_{\rm H}$\,$\lesssim$\,10$^4$\,cm$^{-3}$}) and cold (\mbox{$T_{k}$\,=\,10-20\,K})
envelopes of gas shielded from strong FUV fields
(i.e., negligible abundances of vibrationally excited H$_2$) and the nitrogen chemistry is initiated
by reaction \mbox{N$^+$\,+\,$o$/$p$--H$_2$} \mbox{\citep[e.g.,][]{Dislaire_12,LeGal14}}.}. Protostellar cores such as  \mbox{OMC-2 FIR 4} \citep{Kama13}, the Orion hot core 
\mbox{\citep{Crockett14}}, or \mbox{Orion South} \citep{Tahani16}
do not show  NH emission lines.
Interestingly, bright NH submillimeter and far-IR lines have been  reported in the circumstellar envelope around the eruptive massive binary system \mbox{$\eta$ Carinae} \citep{Morris20,Gull20}.
This unusual nitrogen-rich gas environment 
\citep[\mbox{$N$(NH)\,=\,5$\times$10$^{15}$\,cm$^{-2}$};][]{Gull20} is strongly illuminated
by  X-ray and FUV radiation emitted by the central massive hot evolved stars.
We suspect that much of the NH formation in this complex environment is
driven by reactions of overabundant N atoms with highly vibrationally excited H$_2$.

\section{Summary and conclusion}\label{sec:summary}

Hydrogen abstraction reactions 
\mbox{X\,+\,H$_2$($v$\,=\,0)\,$\rightarrow$\,XH\,+\,H} \mbox{involving}  neutral atoms   such as O, C, S, and N are very endoergic and have substantial  energy barriers.
This implies that even their H$_2$($v\geq1$) vibrational state-dependent reaction rate coefficients rise with an increasing $v$ level and gas temperature. 
 
Due to its high ionization potential, neutral N atoms constitute  the initial reservoir of available gas-phase nitrogen  in  \mbox{FUV-illuminated}  environments.
 We calculated the state-specific rate coefficients of reaction
${\rm N\,(^4\it{S})\,+\,{\rm H_2}\,(v)\,\rightarrow\,{\rm NH\,+\,H}}$ for H$_2$ in \mbox{vibrationally} excited  levels
up to \mbox{$v$\,=\,12}.    
The newly computed rate coefficients imply that reactions 
of N atoms with highly vibrational excited H$_2$ molecules (after \mbox{FUV-pumping}) enhance the formation of NH in   strongly irradiated dense PDRs. For the Orion Bar conditions, we find a total NH column density enhancement of a factor $\sim$25 with respect to models that use the thermal rate coefficient.
We predict that most of the NH column density in the Orion Bar arises from the PDR surface, between 
$A_V$\,$\sim$\,0.5 and 2\,mag, where reactions of N atoms and H$_2$ molecules  in
\mbox{$v$\,$\geq$\,7} vibrational levels  dominate the formation of NH radicals.
Prompted by this result we searched for NH emission lines in the Herschel/HIFI  spectra of the Orion Bar and Horsehead PDRs. Only toward the Bar
we do report a 3$\sigma$ emission feature at the $\sim$974.479\,GHz frequency of the NH 
$N_J = 1_2 - 0_1$ line. This emission level implies a NH column density  of 
about \mbox{10$^{13}$\,cm$^{-2}$},  which can only be matched by  PDR models using the newly computed state-specific rate coefficients.

Owing to very subthermal excitation and endoergic formation, the rare detection of submillimeter NH emission lines seems associated with strongly \mbox{FUV-irradiated} dense gas. In addition to the Orion Bar, another likely candidate is the circumstellar environment around $\eta$ Carinae, where particularly bright NH emission lines have been detected \mbox{\citep{Morris20,Gull20}}.
JWST will soon detect the \mbox{infrared} H$_2$ emission from highly vibrationally excited levels in many \mbox{FUV-irradiated} environments. This will be a unique opportunity to quantify their populations and  role in  interstellar and circumstellar
chemistry. 
 
\begin{acknowledgements}  
We thank Antonio J. C. Varandas (\mbox{N\,+\,H$_2$}), 
George C. Schatz (\mbox{C\,+\,H$_2$}),
Gyorgy Lendvay (\mbox{S\,+\,H$_2$}),
Alex Zanchet  (\mbox{O\,+\,H$_2$}),
Gunnar Nyman (\mbox{N$^+$\,+\,H$_2$}),
Philipe Halvick  (\mbox{C$^+$\,+\,H$_2$}), and
Miguel Gonz\'alez  (\mbox{O$^+$\,+\,H$_2$})
for providing  the potential energy surfaces of the reactions in parenthesis.
We warmly thank David Teyssier for providing the HIFI spectrum of the \mbox{Horsehead PDR} around 974\,GHz and Franck Le Petit and Emeric Bron for useful discussions about vibrationally excited H$_2$. 
We thank the referee for \mbox{concise} but useful suggestions.
JRG thanks the Observatoire de Paris (LERMA) for hosting him when this manuscript was written.
 We thank the Spanish MCINN for funding support under grants
\mbox{PID2019-106110GB-I00} and \mbox{PID2021-122549NB-C21}.

\end{acknowledgements}

%
%

\bibliographystyle{aa}
\bibliography{references}

\begin{appendix}

\section{State-specific rates of reaction 
\mbox{S($^3$P) + H$_2$\,($^1\Sigma_g^+$,v) $\rightarrow$ SH\,(X$^2\Pi_i$) + H}
\label{Sect:Appendix}}

To further support the similar behavior of  \mbox{reaction~(\ref{reaction-general})} when
X is a neutral atom,  we also studied the analogous reaction of H$_2$($v$) molecules
with S($^3P$) atoms. This reaction also presents a late barrier 
(see Fig.~\ref{fig:pes2d}).
\cite{Goicoechea21b} previously reported  quantum wave packet calculations of the
reaction rate coefficients for \mbox{$v=$2} and \mbox{$v=$3}. For $v=3$, the reaction 
becomes exoergic, and the initial vibrational energy of H$_2$ is higher than the energy barrier
 (see Fig.~\ref{fig:reaction_paths}). 
Here we extend the calculations to higher $v$ values using
QCT calculations similar to those described in Sect.~\ref{sub-rates}.

\subsection{QCT reaction rate coefficients up to $v$\,=\,12}

We considered S atoms  in  $^3P_J$ levels, with nine spin-orbit states, with energies
0, 396.640, and 573.64 cm$^{-1}$ for $J$=2, 1, and 0, respectively. We 
introduced the spin-orbit
splitting a posteriori. 

First, we calculated the rate coefficients for each vibrational state of H$_2$($v$) for each adiabatic
electronic state of the H$_2$+S($^3P$) system, neglecting spin-orbit couplings. 
The three adiabatic
states   correlate to  S($^3P_\Lambda$), with $\Lambda$=0, $\pm 1$
  being the projection of the atomic electronic angular momentum. The SH(X$^2\Pi$) state correlates with S($^3P_{\Lambda=\pm 1}$),
  which are degenerate at the collinear geometries considered in Figs.~\ref{fig:reaction_paths} and \ref{fig:pes2d}. The
  S($^3P_{\Lambda= 0}$) state correlates with the SH(A$^2\Sigma^+$) state, which is at 3.85 eV above SH(X$^2\Pi$),
  and therefore its contribution to the reactivity is negligible at the energies considered here.

  The two ground adiabatic states of the SH$_2$ system describing the S($^3$P) + H$_2$($^1\Sigma_g^+$) $\rightarrow$
  SH(X$^2\Pi_i$) + H reaction are the 1$^3A'$ and 1$^3A''$ states, and here we consider the three-dimensional
  potential energy surfaces of  \cite{Maiti-etal:04}. We performed quasi-classical trajectory calculations
  for each of these two adiabatic electronic states, and for H$_2$($v$\,=\,0,1, ... 12). For each electronic vibrational
  states and each temperature, we calculated about 10$^5$ trajectories. The two electronic states present very
  similar rate coefficients, and the rates of the two electronic $^3A'$ and $^3A''$ states were averaged for simplification purposes,
   that is,
  ${\cal K}_{v}(T)=\left[ {\cal K}^{^3A'}_{v}(T)+{\cal K}^{^3A''}_{v}(T)\right]/2$.

Second, considering that these two triply degenerate states correspond to S($^3P_{J=2}$) (five states)
  and to one S($^3P_{J=1}$) (one state) adiabatically, the total rate constant is then given by
  \begin{eqnarray}
    {k}_v(T)= { 5  + e^{- 396.64/K_bT} \over  5  + 3 e^{- 396.64/K_bT} +    e^{- 573.64/K_bT}} {\cal K}_{v}(T)
  .\end{eqnarray}

The obtained rate coefficients  are shown in Fig.~\ref{fig:rates-S}, and the Arrhenius parameters used to fit them are  listed in
\mbox{Table~\ref{tab:rates-parameters-S}} (including our determination
of the thermal rate coefficient). They show a very similar behavior to that of 
reaction \mbox{N($^4S$)\,+\,H$_2$($v$)} discussed in the main text. In particular, the rate coefficients show
a threshold even for $v=3$, for which the reaction is exoergic and the vibrational energy is above the reaction barrier. This implies rate coefficients that have
a positive dependence with temperature, even for vibrational excited
H$_2$ levels. On the other hand, the analogous hydrogen abstraction reaction with S$^+$ ions,
reaction \mbox{S$^+(^4S)$\,+\,H$_2$($v$)\,$\rightarrow$\,SH$^+$\,+\,H$_2$},  shows the typical behavior of an exothermic reaction for \mbox{$v$\,$\geq$\,2}, in other words a rate
coefficient that is  nearly independent of temperature (see Fig.~\ref{fig:compa_rates}).\\

\begin{figure}[ht]
\centering   
\includegraphics[scale=0.6, angle=0]{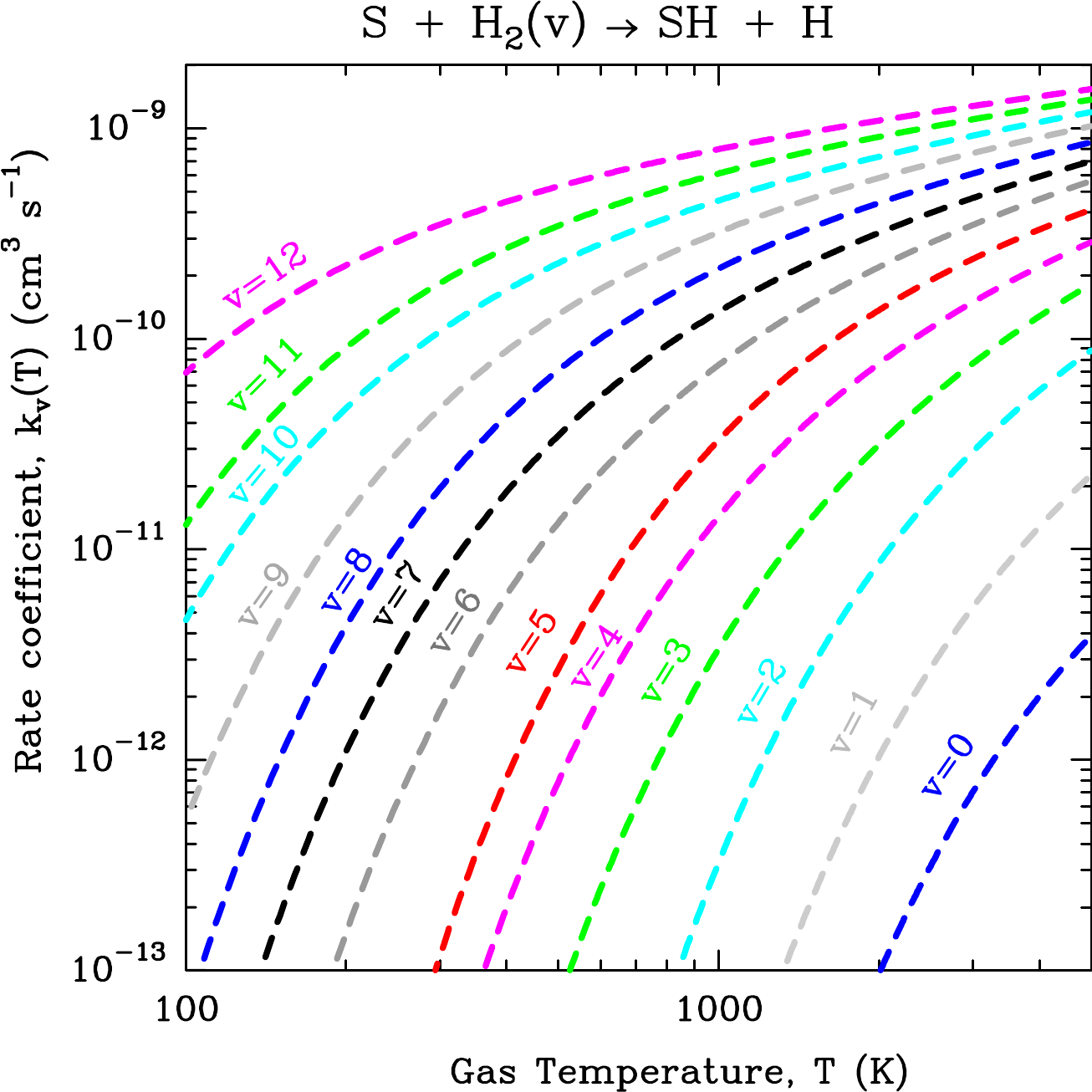}
\caption{QCT state-specific rate coefficients of reaction \mbox{S($^3P$) + H$_2$\,($^1\Sigma_g^+$,v) $\rightarrow$ SH\,(X$^2\Pi_i$) + H} computed in this work.}
\label{fig:rates-S}
\end{figure}

\begin{table}[h]
\centering
\caption{H$_2$ vibrational energies ($E_v$) and Arrhenius fit parameters,
 $k_v(T) = \alpha \,(T/300)^{\beta} e^{-\gamma/T}$, of the QCT state-specific rate coefficients  calculated 
 in this study for
 reaction \mbox{S($^3P$)\,+\,H$_2$($v$)\,$\rightarrow$\,SH\,+\,H}. }
\begin{tabular}{c  c r c r}
  \hline\hline
  $v$ & $E_v$ (eV) & $\alpha$ (cm$^3$\,s$^{-1}$) & $\beta$ & $\gamma$ (K)  \\
  \hline
           0 &  0.270    &   0.052\,$\times$\,10$^{-10}$  &  0.636  &  10390.9 \\
           1 &  0.784    &   0.140\,$\times$\,10$^{-10}$  &  0.738  &  8000.0 \\
           2 &  1.270    &   0.363\,$\times$\,10$^{-10}$  &  0.712  &  5572.2 \\
           3 &  1.727    &   0.337\,$\times$\,10$^{-10}$  &  0.835  &  3305.9 \\
           4 &  2.156    &   0.766\,$\times$\,10$^{-10}$  &  0.646  &  2454.9 \\
           5 &  2.557    &   1.590\,$\times$\,10$^{-10}$  &  0.492  &  2163.0 \\
           6 &  2.931    &   1.362\,$\times$\,10$^{-10}$  &  0.598  &  1312.4 \\
           7 &  3.275    &   1.914\,$\times$\,10$^{-10}$  &  0.531  &  989.3   \\
           8 &  3.599    &   2.718\,$\times$\,10$^{-10}$  &  0.466  &  790.3 \\
           9 &  3.884    &   3.567\,$\times$\,10$^{-10}$  &  0.419  &   608.7   \\
           10 &  4.134   &   4.227\,$\times$\,10$^{-10}$  &  0.399  &   407.6\\
           11 &  4.349   &   5.773\,$\times$\,10$^{-10}$  &  0.332  &   342.4  \\
           12 &  4.523   &   6.632\,$\times$\,10$^{-10}$  &  0.313  &    191.5    \\\hline
              & thermal  &  4.000\,$\times$\,10$^{-10}$ &        &  15500.0 \\  
\hline
 \end{tabular}
\label{tab:rates-parameters-S} 
\end{table}

\clearpage

\subsection{Impact on the SH abundance in the Orion Bar}

Figure~\ref{fig:mods-S} shows the impact of the  state-specific rate coefficients of reaction 
\mbox{S\,($^3P$) + H$_2$\,($v$) $\rightarrow$ SH\,+\,H},
computed here up to $v$\,=\,12, in a PDR model of the Orion Bar. The pink continuous curve shows the SH abundance profile  using these rate coefficients. The pink dashed curve is for a model that
uses the thermal rate coefficient.
Despite these state-specific rate coefficients showing the same fundamental behavior as
those of reaction \mbox{N($^4S$)\,+\,H$_2$($v$)\,$\rightarrow$\,NH\,+\,H}, the impact on the formation of SH radicals is very minor (an increase of $\sim$\,1\% in column density).
This difference arises from the fact that S$^+$ ions, and not S atoms, are the dominant gas-phase sulfur reservoir in the PDR surface layers where highly vibrationally excited H$_2$ exist. 

\citet{Goicoechea21b} present a detailed study of the chemistry of sulfur bearing hydrides in the Orion Bar. The SH abundance profile shown in \mbox{Fig.~\ref{fig:mods-S}} is very similar to their
Fig.~12 (\mbox{implementing} quantum rate coefficients up to \mbox{$v$\,=\,3}
for reaction  \mbox{S($^3P$)\,+\,H$_2$($v$)\,$\rightarrow$\,SH\,+\,H}).

\begin{figure}[h]
\centering   
\includegraphics[scale=0.42, angle=0]{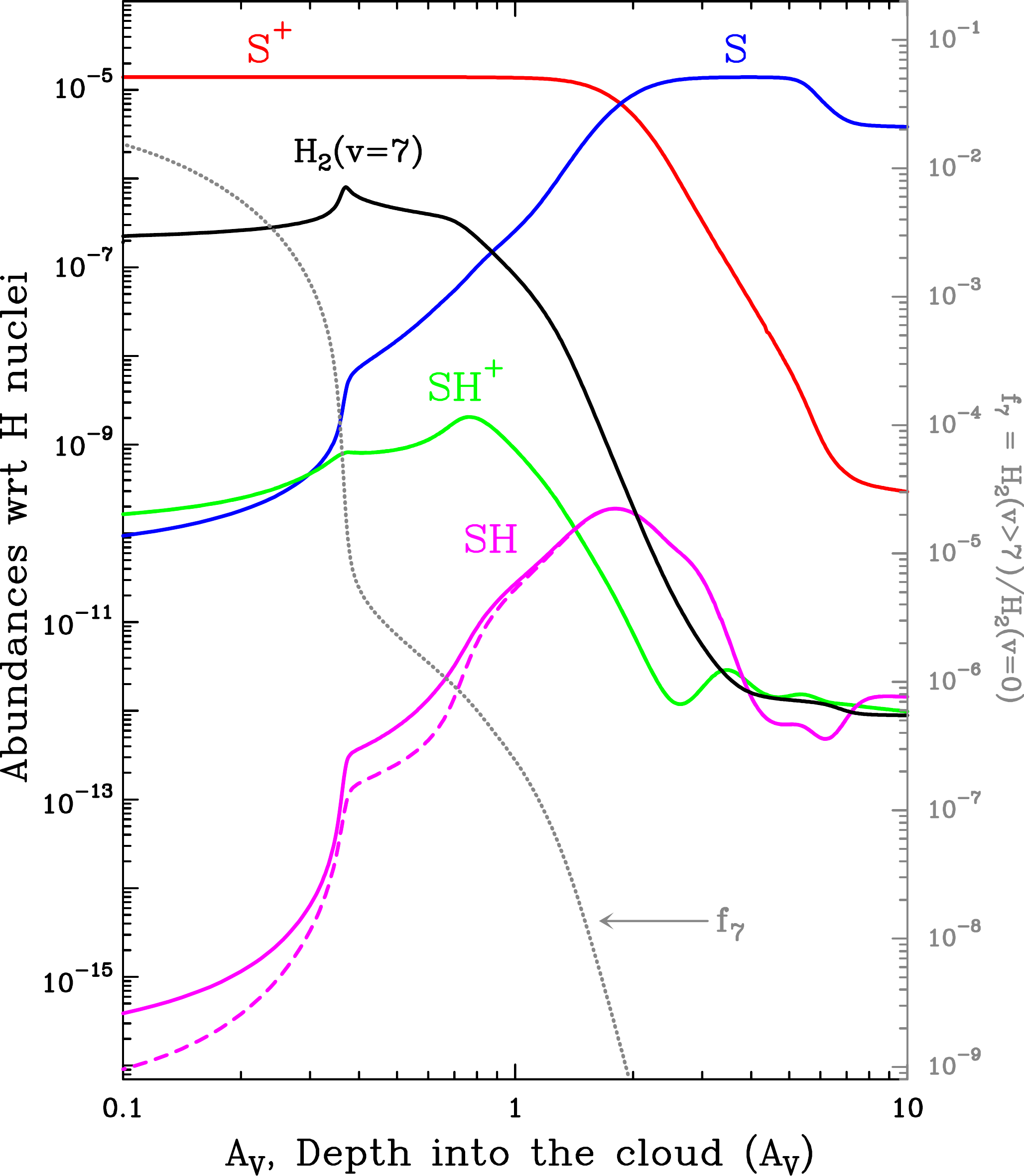}
\caption{Isobaric PDR model of the Orion Bar  (\mbox{$G_0$\,$\simeq$2$\times$10$^4$},  
\mbox{$P_{\rm th}/k_{\rm B}$\,=\,2$\times$10$^8$\,cm$^{-3}$\,K}) showing
 abundance profiles with respect to H nuclei. The gray dotted curve shows $f_7$, the fraction of \mbox{H$_2$($v>7$)} with respect to the ground (right axis gray scale).
Solid curves refer to a model using state-specific reaction rates
for reaction~\mbox{S\,($^3P$) + H$_2$\,($v$) $\rightarrow$ SH\, + H,}  whereas dashed curves refer to a model using the thermal rate.}
\label{fig:mods-S}
\end{figure}

\end{appendix}

\end{document}